\documentclass{jaa}

\usepackage{graphicx}
\usepackage{xspace}	

\usepackage{threeparttable}
\usepackage{amsmath}
\usepackage[utf8]{inputenc}
\usepackage{longtable}
\usepackage{graphicx}
\usepackage{natbib}
\usepackage{multirow}
\usepackage[Table,xcdraw]{xcolor}
\usepackage{rotating}
\newcommand{\apj}{ApJ}

\newcommand{\mnras}{MNRAS}

\newcommand{\physrep}{Physics Reports}
\newcommand{\nat}{Nature}
\newcommand{\apjl}{ApJL}
\newcommand{\apjs}{ApJS}

\newcommand{\aap}{Astronomy and Astrophysics}

\newcommand{\aj}{AJ}

\newcommand{\pasp}{Publications of the Astronomical Society of the Pacific}

\newcommand{\fermi}{{\em Fermi}\xspace}

\newcommand{\keV}{{\rm keV}\xspace}

\newcommand{\swift}{{\em Swift}\xspace}
\newcommand{\tninty}{{$T_{\rm 90}$}\xspace}

\newcommand{\sw}[1]{\texttt{#1}}

\usepackage{hyperref}
\hypersetup{colorlinks=true,
    breaklinks=true,
    urlcolor= black,
    linkcolor= blue,
    bookmarksopen=false,
    filecolor=black,
    citecolor=blue,
    linkbordercolor=blue
}

\begin{document}\sloppy
%
\title{Photometric studies on the host galaxies of gamma-ray bursts using 3.6m Devasthal Optical Telescope}

\author{Rahul Gupta\textsuperscript{1,2,*}, Shashi Bhushan Pandey\textsuperscript{1}, Amit Kumar\textsuperscript{1,3}, Amar Aryan\textsuperscript{1,2}, Amit Kumar Ror\textsuperscript{1}, Saurabh Sharma\textsuperscript{1}, Kuntal Misra\textsuperscript{1}, A. J. Castro-Tirado\textsuperscript{4,5}, and Sugriva Nath Tiwari\textsuperscript{2}}
\affilOne{\textsuperscript{1}Aryabhatta Research Institute of Observational Sciences (ARIES), Manora Peak, Nainital-263002, India.\\}
\affilTwo{\textsuperscript{2}Department of Physics, Deen Dayal Upadhyaya Gorakhpur University, Gorakhpur-273009, India.\\}
\affilThree{\textsuperscript{3}School of Studies in Physics and Astrophysics, Pt. Ravishankar Shukla University, Chattisgarh 492010, India.\\}
\affilFour{\textsuperscript{4}Instituto de Astrof\'isica de Andaluc\'ia (IAA-CSIC), Glorieta de la Astronom\'ia s/n, E-18008, Granada, Spain.\\}
\affilFive{\textsuperscript{5}Departamento de Ingenier\'ia de Sistemas y Autom\'atica, Escuela de Ingenier\'ias, Universidad de M\'alaga, C\/. Dr. Ortiz Ramos s\/n, 29071 M\'alaga, Spain.\\}
\twocolumn[{

\maketitle

\corres{rahulbhu.c157@gmail.com, rahul@aries.res.in}

\msinfo{6 December 2021}{27 May 2022}

\begin{abstract}

In this article, we present multi-band photometric observations and analysis of the host galaxies for a sample of five interesting gamma-ray bursts (GRBs) observed using the 3.6m Devasthal Optical Telescope (DOT) and the back-end instruments. The host galaxy observations of GRBs provide unique opportunities to estimate the stellar mass, ages, star-formation rates, and other vital properties of the burst environments and hence progenitors. We performed a detailed spectral energy distribution (SED) modeling of the five host galaxies using an advanced tool called \sw{Prospector}, a stellar population synthesis model. Furthermore, we compared the results with a larger sample of well-studied host galaxies of GRBs, supernovae, and normal star-forming galaxies. Our SED modeling suggests that GRB 130603B, GRB 140102A, GRB 190829A, and GRB 200826A have massive host galaxies with high star formation rates (SFRs). On the other hand, a supernovae-connected GRB 030329 has a rare low-mass galaxy with a low star formation rate. We also find that GRB 190829A has the highest (in our sample) amount of visual dust extinction and gas in its local environment of the host, suggesting that the observed very high energy emission from this burst might have a unique local environment. Broadly, the five  GRBs in our sample satisfy the typical correlations between host galaxies parameters and these physical parameters are more common to normal star-forming galaxies at the high-redshift Universe. Our results also demonstrate the capabilities of 3.6m DOT and the back-end instruments for the deeper photometric studies of the host galaxies of energetic transients such as GRBs, supernovae, and other transients in the long run.  

\end{abstract}


}]


\doinum{https://doi.org/10.1007/s12036-022-09865-0}
\volnum{43}
\year{2022}
\pgrange{1-}
\artcitid{\#\#\#\#}
\lp{17}
\setcounter{page}{1}

\section{Introduction}

Gamma-ray bursts (GRBs) are the most luminous and fascinating sources observed in the Universe since the Big bang. Their unique characteristic properties provide an excellent opportunity to study compact binaries, the evolution of massive stars, and extreme physical phenomenon out to very large distances \citep{2015PhR...561....1K}. GRBs are supposed to be originated from relativistic jets launched either due to the merger of two compact objects producing short GRBs (\tninty $\leq$ 2 sec) or due to the birth of a stellar-mass black hole or a rapidly rotating magnetized neutron star during the core-collapse of massive stars giving rise to long GRBs (\tninty $>$ 2 sec). 
However, the origin of a few short bursts (e.g., GRB 090426 and GRB 200826A) from the collapse of massive stars and the origin of a few long bursts (e.g., GRB 060614 and GRB 211211A) from the binary merger challenges our current understanding of the nature of possible progenitors of GRBs. These examples suggest that at least some of the short GRBs might be originated from collapsars \citep{2009A&A...507L..45A, 2021NatAs.tmp..142A, 2021NatAs...5..911Z} and some of the long GRBs might be originated from mergers \citep{2006Natur.444.1050D, 2022arXiv220903363T}. Followed by the prompt emission and longer-lasting multi-wavelength afterglow phases \citep{2004RvMP...76.1143P, 2019Natur.575..455M}, late-time observations of the host galaxies are of crucial importance to examine the burst environment and, in turn, about the possible progenitors.   

GRBs can be used to study the galaxies both at high (the most distant, GRB 090423 with spectroscopic $z \sim$ 8.2) and low (nearest, GRB 170817A $z \sim$ 0.0097) redshifts due to their intrinsic brightness (much higher signal to noise ratio; \citealt{2012A&A...542A.103B}). The GRB host galaxy characteristics of long and short GRBs are largely different such as morphology, stellar population, offsets, etc. but share common properties too for a fraction of observed populations. These observed characteristics are likely associated with the physical conditions surrounding possible progenitors producing GRBs. Long GRBs are generally localized in active star-forming and young stellar population dwarf galaxies. Since long GRBs are likely to be related to the death of massive stars, they are widely cited as robust and potentially unbiased tracers of the star formation and metallicity history of the Universe up to z $\sim$ eight \citep{2009ApJ...691..182S}. Host galaxy observations of long GRBs suggest that they preferentially occur in low-metallicity galaxies \citep{2011MNRAS.414.1263M}. On the other hand, short GRBs are expected in any type of galaxy associated with an old stellar population \citep{2009ApJ...690..231B}. Their locations relative to their host centers 
have a median physical offset of about five kpc, which is about four times larger than the median offset for long bursts \citep{2013ApJ...776...18F}. Therefore, host parameters can constrain the nature of GRBs' possible progenitors and environments.

In the pre-\swift era (before 2004), there were few bursts with measured redshifts. In this era, the host galaxies were intensely studied once the redshift was known to be low ($z$ $\le$ 0.3). In \swift era, the number of GRBs with measured redshift values increases but still $\sim$ 25 \% of the localized ones and could still be biased against dusty events\footnote{\url{https://www.mpe.mpg.de/~jcg/grbgen.html}}. \cite{2009ApJ...691..182S} studied the host galaxy properties for a large sample of GRBs hosts and suggested that GRB hosts are similar to normal star-forming galaxies in both the nearby and the distant universe. \cite{2016ApJ...817....7P, 2016ApJ...817....8P} examined an unbiased sample \footnote{introduced as the \swift Gamma-Ray Burst Host Galaxy Legacy Survey} of the host galaxies of long GRBs (mainly photometric) and proposed that the dusty bursts are generally found in massive host galaxies. It gives a clue that the massive galaxies (star-forming) are typical and homogeneously dusty at higher redshift. On the other hand, low-mass galaxies (star-forming) have a small amount of dust in their interstellar medium (to some level). Also, \cite{2010MNRAS.405...57S, 2018A&A...617A.105J} presented comparative studies of the host galaxies of GRBs and compared their properties with those of core-collapse supernovae (CCSNe). More recently, \cite{2021MNRAS.503.3931T} presented a comprehensive study of a large sample of core-collapse supernova (CCSN) host galaxies and compared with the host galaxies of the nearest long GRBs and superluminous supernova (SLSN) and found a hint that host-galaxy mass or specific star-formation rate is more fundamental in driving the preference for SLSNe and long GRBs in unusual galaxy environments.

Deeper optical photometric follow-up observations of energetic transients such as afterglows of long and short GRBs are frequently carried out \citep{2021RMxAC.53....127, 2020GCN.29148....1P, 2021arXiv211111795G, 2021GCN.29490....1G, 2021GCN.31299....1G, 2022MNRAS.511.1694G, 2022MNRAS.tmp.1031K} using the recently commissioned largest Indian optical telescope, i.e., 3.6m DOT situated at Devasthal observatory of Aryabhatta Research Institute of Observational Sciences (ARIES) Nainital and the back-end instruments \citep{2018BSRSL..87...42P, 2018BSRSL..87...58O, 2021arXiv211113018K}. Observations of galaxies and other objects of low surface brightness are also carried out using the 3.6m DOT \citep{2021arXiv211113018K, 2021arXiv211111796P}. In this work, we performed the spectral energy distribution modeling of a sample of five host galaxies of GRBs observed by the 3.6m DOT/back-ends and compared the results with other well-studied samples of host galaxies. We observed the host galaxies of these five bursts subject to the availability of the observing time of the CCD IMAGER and clear sky conditions (see section \ref{sample}). This work demonstrates the capabilities of deep follow-up observations of such faint and distant hosts of explosive transients using the 3.6m DOT. We have arranged this article in the following sections. In section \ref{sample}, we present our host galaxies sample (with brief details about each burst) and their multi-band photometric observations taken with 3.6m DOT. In section \ref{sedmodeling}, we present the host galaxy spectral energy distribution modeling of our photometric data along those obtained from literature using \sw{Prospector} software (version 1.1.0). In section \ref{results}, we have given the SED modeling results and comparison with other known host galaxies. Finally, in section \ref{Summary and Conclusion}, we have presented the summary and conclusion of the present work. Throughout this paper, we consider following cosmological values: the Hubble parameter $\rm H_{0}$ = 70 km $\rm s^{-1}$ $\rm Mpc^{-1}$, density parameters $\rm \Omega_{\Lambda}= 0.73$, and $\rm \Omega_m= 0.27$.

\section{Sample of Host galaxies, observations with the 3.6m DOT}
\label{sample}

\begin{figure}[!ht]
\includegraphics[scale=0.35]{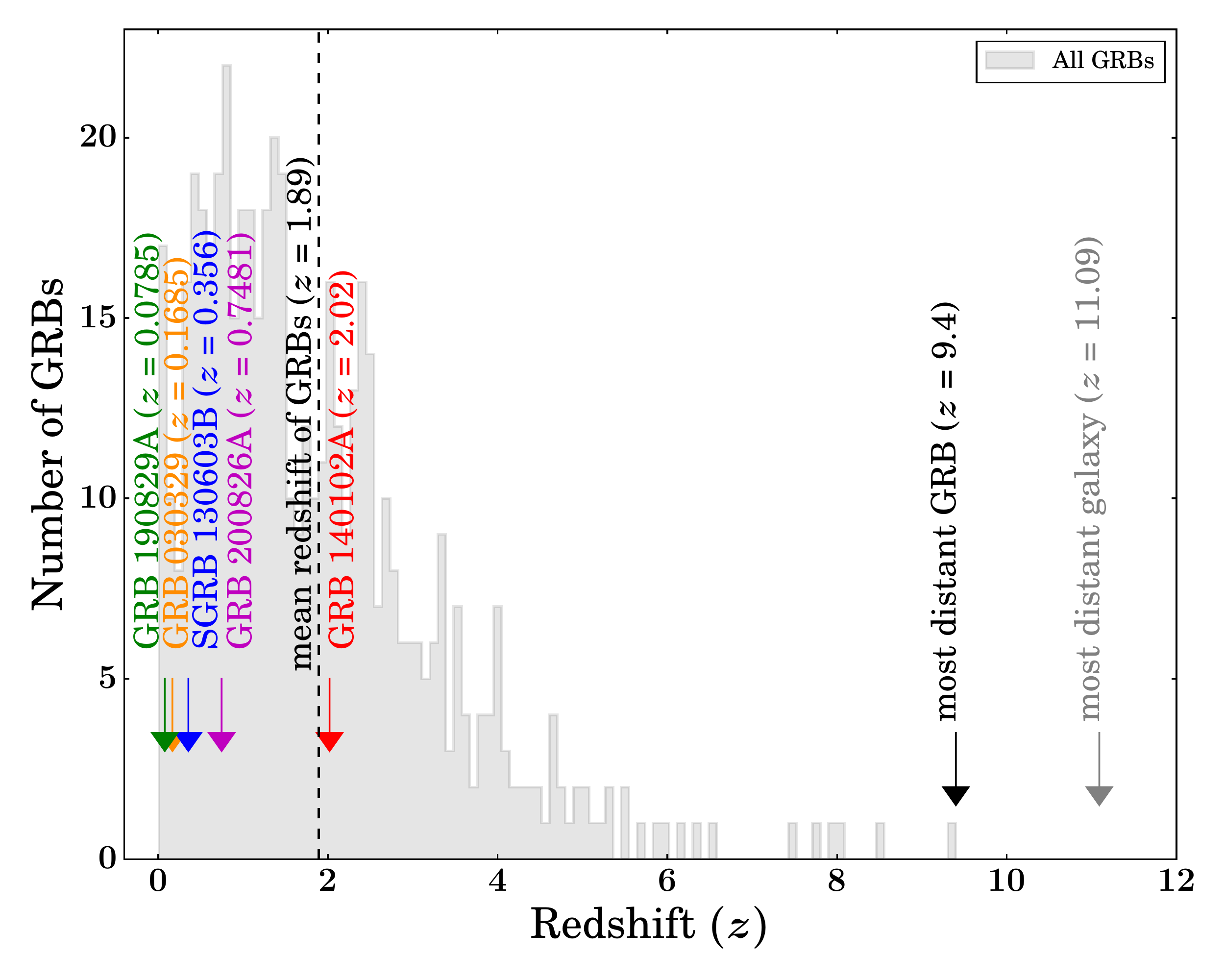}
\caption{The redshift distribution for all the GRBs with a measured redshift value till January 2021 (shown with grey color). The position of each burst of our sample is also shown. The vertical black dashed line represents the mean value of the redshift for all the GRBs with a measured redshift.} 
\label{redshift}
\end{figure}

In this section, we provide details of multi-band photometric observations of the host galaxies of the sample (five bursts with peculiar features between redshifts 0.0758 to 2.02, see details below), using India's largest 3.6m DOT telescope and back-end instruments like 4Kx4K CCD Imager \citep{2018BSRSL..87...42P} and the TANSPEC \citep{2018BSRSL..87...58O}. Deeper optical observations of the host galaxies of GRB 030329, GRB 130603B, GRB 140102A, and GRB 190829A in several optical filters ($B, V, R, I$) were obtained using the 4Kx4K CCD Imager. Details about observations of each of these four host galaxies are described below in respective sub-sections. We performed the optical photometric data analysis for the host galaxies observations using IRAF/DAOPHOT using methods described in \citep{2021arXiv211113018K, 2021MNRAS.505.4086G, 2019MNRAS.485.5294P}. In the case of GRB 200826A, photometric optical-NIR observations in $I, J$, and $K$ filters were performed using the TANSPEC \citep{2018BSRSL..87...58O}, and the detail about the data reduction are described in the respective subsection below. A photometric observation log for each burst of our sample is given in Table \ref{tab:observationslog} of the appendix. The redshift distribution of all the GRBs with a measured redshift up to January 2021 (data obtained from GRBweb catalog page\footnote{\url{https://user-web.icecube.wisc.edu/~grbweb_public/Summary_table.html}} provided by P. Coppin), along with those discussed presently are shown in Fig. \ref{redshift}.

\subsection{GRB 030329 (associated SN 2003dh):}

GRB 030329 was triggered by many detectors on-board the High Energy Transient Explorer (HETE-2) mission at 11:37:14.67 UT on 2003 March 29. The prompt emission light curve of this GRB consists of two merging emission pulses with a total duration of $\sim$ 25 s in 30-400 \keV energy band. Later on, a multiwavelength follow-up observations campaign of GRB 030329 revealed the discovery of optical \citep{2003GCN..1985....1P}, X-ray \citep{2003GCN..1996....1M}, and radio counterparts \citep{2003GCN..2014....1B}. \cite{2003GCN..2020....1G} measured the redshift of the burst ($z$) = 0.1685 using the early spectroscopy observations taken with the Very Large Telescope (VLT). Furthermore, a late-time bump in the optical light curve along with spectroscopic observations confirms detection of associated broad-line type Ic supernova \citep{2003Natur.423..847H}, establishing the relationship between the afterglow of long GRBs and supernovae. \cite{2008MNRAS.387.1227O} utilizes the spectral evolutionary models to constrain the progenitor and suggested collapsar scenario for the progenitor of GRB 030329/SN 2003dh.

We performed the host galaxy observations of GRB 030329 using 4K $\times$ 4K Imager mounted on the axial port 3.6m DOT in March 2017. Multiple frames with an exposure time of 600 s were taken in R and V filers. The host galaxy of GRB 030329/SN 2003dh is clearly detected in both filters. A finding chart taken in the R filer is shown in the upper-left panel of Fig. \ref{host_findingchart}.

\subsection{Short GRB 130603B (associated kilonova emission):}

This burst was detected by Burst Alert Telescope (BAT) on-board \swift mission at 15:49:14 UT on 03$^{rd}$ June 2013 at the position RA = 172.209, DEC = +17.045 degree (2000) with an uncertainty of three arcmin. The prompt emission BAT light curve consists of the fast-rising and exponential decay (FRED) like single structure with \tninty duration of 0.18 $\pm$ 0.02 s (in 15-350 \keV), classifying this burst as a short-duration GRB \citep{2013GCN.14741....1B}. \cite{2013GCN.14744....1T} obtained the afterglow spectrum using 10.4m GTC and reported the redshift of the burst $z$ = 0.356. Later on, the late time near-infrared (NIR) observations reveal the detection of the kilonova emission (the first known case) accompanying with short GRB 130603B, supporting the merger origin of short GRBs \citep{2013Natur.500..547T}. \cite{2014A&A...563A..62D} studied the environment and proposed that the explosion site of this burst is not similar to those seen in the case of long GRBs.

\begin{figure*}[ht!]
\centering
\includegraphics[angle=0,scale=0.282]{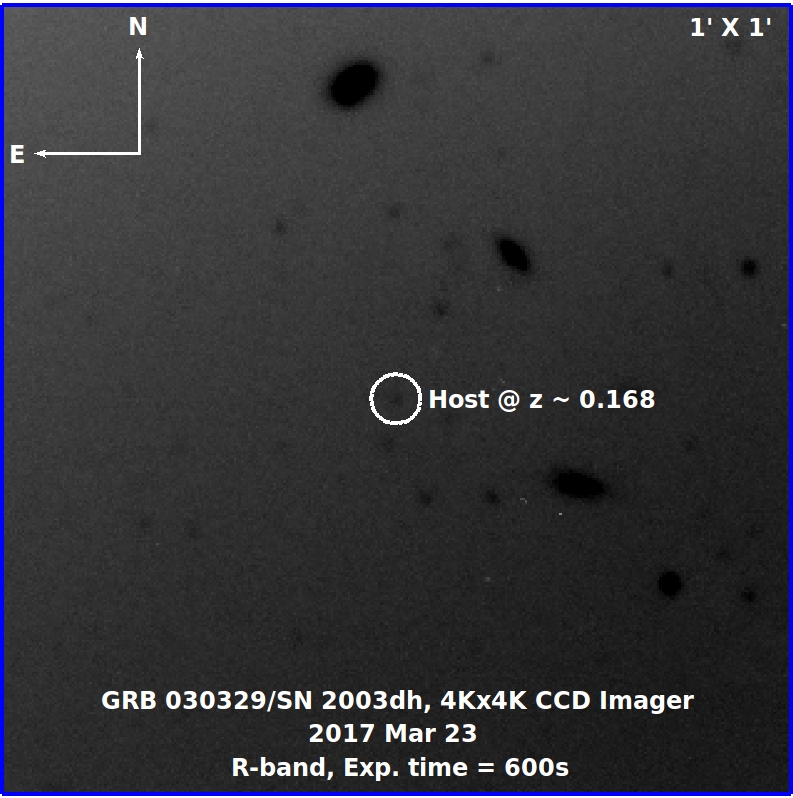}
\includegraphics[angle=0,scale=0.285]{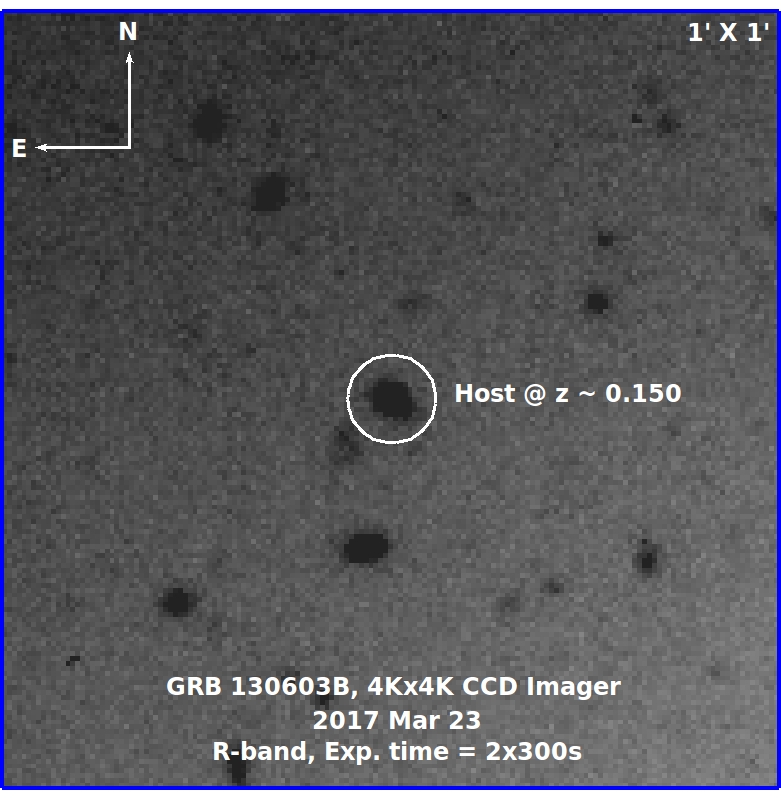}
\includegraphics[angle=0,scale=0.28]{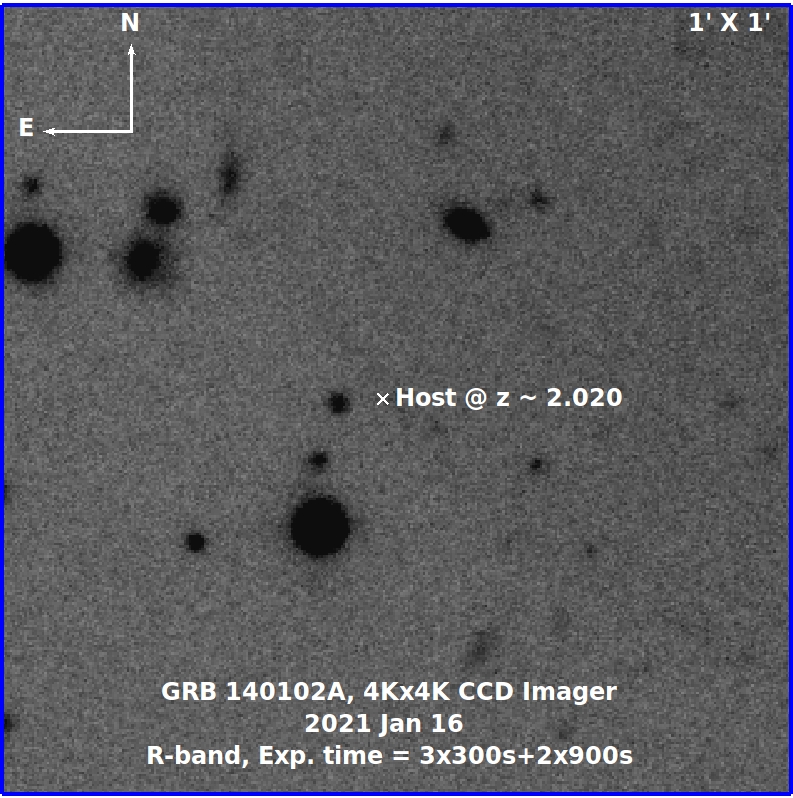}
\includegraphics[angle=0,scale=0.285]{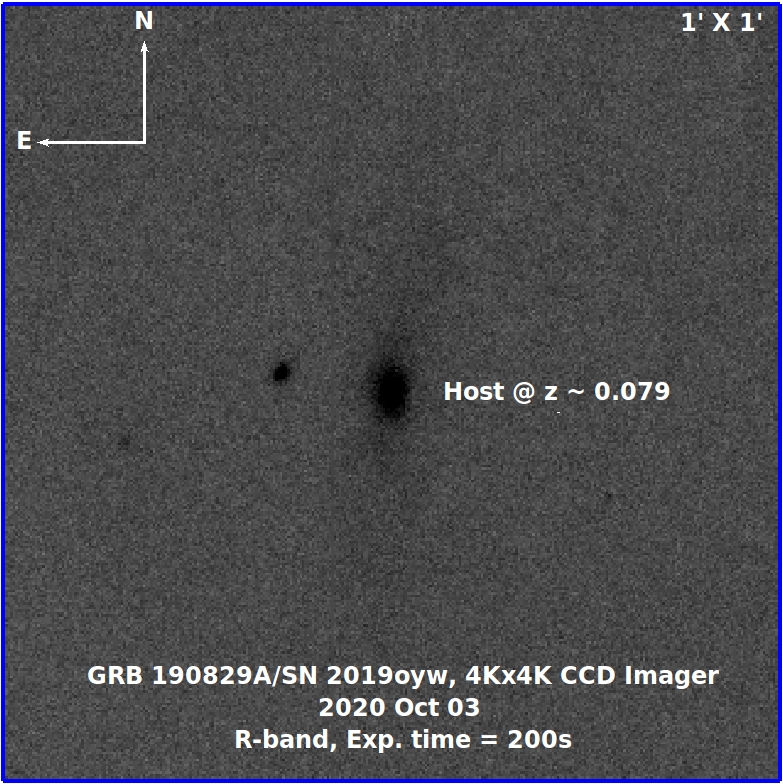}
\caption{The R-band finding charts of the host galaxies of GRB 030329 (top left), GRB 130603B (top right), GRB 140102A (bottom left), and GRB 190829A (bottom right) were obtained using 4K$\times$4K CCD Imager mounted on the 3.6m DOT \citep{2018BSRSL..87...42P, 2021arXiv211113018K}. The position of the host galaxies in the frames is marked with circles.}
\label{host_findingchart}
\end{figure*}

We observed the host galaxy of short GRB 130603B using 3.6m DOT in B and R filters with an exposure time of 2x300 s in each on $23^{rd}$ March 2017. We detected a bright galaxy (see the upper-right panel of Fig. \ref{host_findingchart}) in both the filters. 

\subsection{GRB 140102A (early reverse shock emission):}

GRB 140102A was jointly detected by \fermi (by both Gamma-ray Burst Monitor (GBM) and Large Area Telescope (LAT)) and \swift BAT detectors. We carried out an early follow-up of the optical afterglow of GRB 140102A reveals the rarely observed reverse shock signature in the light curve. We also constrain the redshift of the burst $z$ = 2.02 using joint X-ray and optical SED. We calculated the magnetization parameter using the afterglow modeling of GRB 140102A and suggested that the jet composition could be dominant with a moderately magnetized outflow, in this case, \citep{2021MNRAS.505.4086G}. 
Furthermore, we carried out the host galaxy observations of GRB 140102A in the R filter (with a total exposure time of 45 minutes) using 3.6m DOT on 16 January 2021. However, we could not detect the host galaxy (see the lower-left panel of Fig. \ref{host_findingchart}), but we could constrain the deep limiting magnitude. 

\subsection{GRB 190829A (nearest Very high energy (VHE) burst):}

GRB 190829A was detected by \fermi GBM (at 19:55:53 UT), and \swift BAT (at 19:56:44.60 UT) on $29^{th}$ August 2019. \cite{2019ATel13052....1D} reported the detection of very high energy (V.H.E.) emission from the source using H.E.S.S. observations. We studied the prompt emission characteristics of the two-episodic low-luminous GRB 190829A, and found that the first episode is an Amati outlier, showing the peculiar nature of the prompt emission \citep{2020ApJ...898...42C, 2021RMxAC..53..113G}. Furthermore, we calculated the redshift of the burst $z$ = 0.0785 using 10.4m GTC spectroscopic observations, making the event the nearest V.H.E. detected GRB along with the detection of associated supernova \citep{2021A&A...646A..50H}. However, it is still an unsolved question that V.H.E. emission is originated due to their environment or its emission mechanism \citep{2020A&A...633A..68D}.

The associated host galaxy of GRB 190829A is a significantly bright SDSS galaxy (SDSS J025810.28-085719.2). We observed this galaxy using 3.6m DOT in B, R, and I filters on 03$^{rd}$ October 2020. The host galaxy is clearly detected in each filter of our observations. A finding chart for the host galaxy of GRB 190829A is shown in the lower-right panel of Fig. \ref{host_findingchart}.

\subsection{GRB 200826A (shortest long burst):}
GRB 200826A was detected by \fermi GBM at 04:29:52.57 UT on 26 August 2020 with a \tninty duration of 1.14 s in the GBM 50-300 \keV energy range \citep{2020GCN.28287....1M}. Late-time optical follow-up observations revealed the bump in the light curve, consistent with the supernova emission. Although the prompt properties of the burst are typical to those of short GRB, late-time follow-up observations confirm a collapsar origin \citep{2021NatAs.tmp..142A}. \cite{2021NatAs.tmp..142A} suggested that the burst is the shortest long burst with SN bump and lie on the brink between a successful and a failed collapsar.

\begin{figure}[ht!]
\includegraphics[angle=0,scale=0.35]{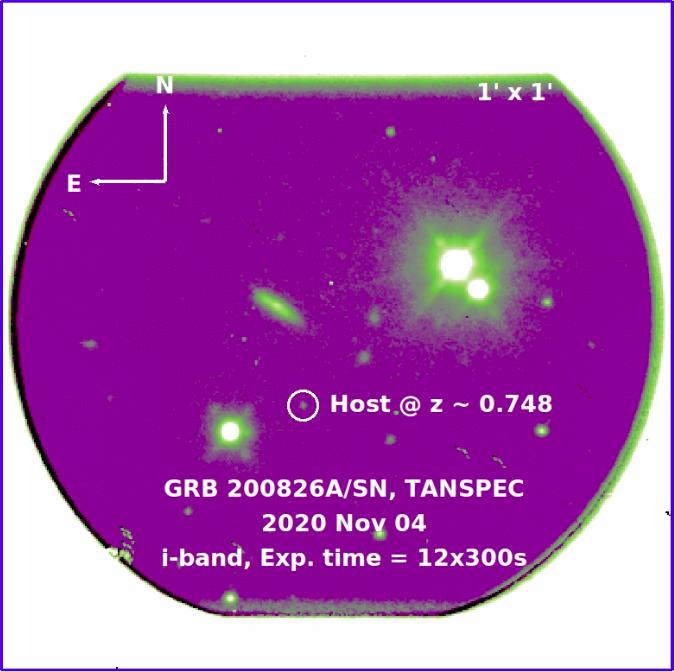}
\caption{The i-band finding chart of the host galaxy of GRB 200826A/SN was obtained using the TANSPEC \citep{2018BSRSL..87...58O}. The position of the host galaxy in the frame is marked with a circle.}
\label{host_findingchart_tanspec}
\end{figure}

We have obtained the optical and NIR photometric data of the host galaxy of GRB 200826A using the TIFR-ARIES Near-Infrared Spectrometer \citep[TANSPEC;][]{2018BSRSL..87...58O} mounted on the 3.6m DOT, Nainital, India during the nights of 2020 November. TANSPEC is a unique instrument that provides simultaneous wavelength coverage from 0.5-2.5 $\mu$m in imaging and spectroscopic modes. We have given exposures of one hour, 35 minutes, 35 minutes in I (4$^{th}$ November 2020), J (4$^{th}$ November 2020), K (11$^{th}$ November 2020) bands, respectively. In I band, 12 frames of 5 min exposure, whereas, in $J$ and $K$ bands, three sets of $20\times5$ sec exposure at seven dithered positions (total of 35 minutes in $J$ and $K$ bands) were taken with TANSPEC. We have used standard data reduction procedures for the image cleaning, photometry, and astrometry \citep[for details, see ][]{2020MNRAS.498.2309S}. The host galaxy is detected in the I band as $22.71 \pm 0.10$ mag, and there was an upper limit of J $>$ 20.56 mag and K $>$ 19.55 mag. A finding chart of the host galaxy of GRB 200826A taken with TANSPEC is shown in Fig. \ref{host_findingchart_tanspec}. 

\section{Panchromatic SED modeling}
\label{sedmodeling}

In our previous studies \citep{2019MNRAS.485.5294P, 2021MNRAS.505.4086G}, \sw{LePHARE} software is used for the modeling of the host galaxies, and it suffers from a major limitation, i.e., using only chi-square statistics to the best fit solution. The results of \sw{LePHARE} software are primarily affected by degeneracy among parameters as it could not provide the posterior distributions. Therefore, in this work, we utilized an advanced software called \sw{Prospector} (version 1.1.0) for SED fitting to constrain the host galaxies properties of our sample of five hosts. \sw{Prospector} \citep{2017ApJ...837..170L, 2021ApJS..254...22J} software (stellar population modeling code) for modeling the SEDs using the measured photometric magnitudes values of the host galaxies. \sw{Prospector} utilizes a library of Flexible Stellar Population Synthesis models \citep{2009ApJ...699..486C}. It is advanced software that determines the best-fit solution to the SED model fitting using \sw{Dynesty} (implements dynamic nested sampling algorithm) and produces the posterior distributions for the model parameters\footnote{\url{ https://github.com/bd-j/prospector}}. The posterior distributions are useful to verify the degeneracy between the model parameters. We performed the SED fitting to photometric data for each of the host galaxies of our sample at their respective fixed redshift values. We have used the \sw{parametric$\_$sfh} model to calculate the stellar population properties such as stellar mass formed ( $M_\star$, in units of solar mass), stellar metallicity ($\log Z/Z_\odot$), age of the galaxy ($t$), rest-frame dust attenuation for a foreground screen in mags (A$_{\rm V}$), and star formation timescale ($\tau$) for an exponentially declining star formation history (SFH). We have set these host galaxy model parameters free to determine the posterior distribution and consider uniform priors across the allowed parameter space within Flexible Stellar Population Synthesis models. We have fixed the maximum values of the prior of the age of the galaxies corresponding to the age of the Universe at their respective measured redshifts values. For the host SED modeling using stellar population models, we consider Milky Way extinction law \citep{1989ApJ...345..245C} and Chabrier initial mass function \citep{2003PASP..115..763C}. We calculated the star formation rate using the following equation taken from \cite{2020ApJ...904...52N}:

\begin{equation*}
\text{SFR}(t) = M\times \Big[\int_0^t{te^{-t/\tau} dt}\Big]^{-1} \times te^{-t/\tau} \hspace{1.75cm} (1) 
\end{equation*} 

\section{Results}
\label{results}

\subsection{SED modeling}

This section presents the results of the host galaxies modeling of our sample. We corrected the values of the observed magnitudes (AB system) for the foreground galactic extinction for each galaxy following \cite{2011ApJ...737..103S} and used them as input to \sw{Prospector} for the SED modeling of galaxies. The best-fit SEDs are shown in Fig. \ref{sed} (corner plots are shown in Fig. \ref{corner} of the appendix), and their results thus obtained are tabulated in Table \ref{tab:sedresults}.

\begin{figure*}[!ht]
\centering
\includegraphics[width=.9\columnwidth]{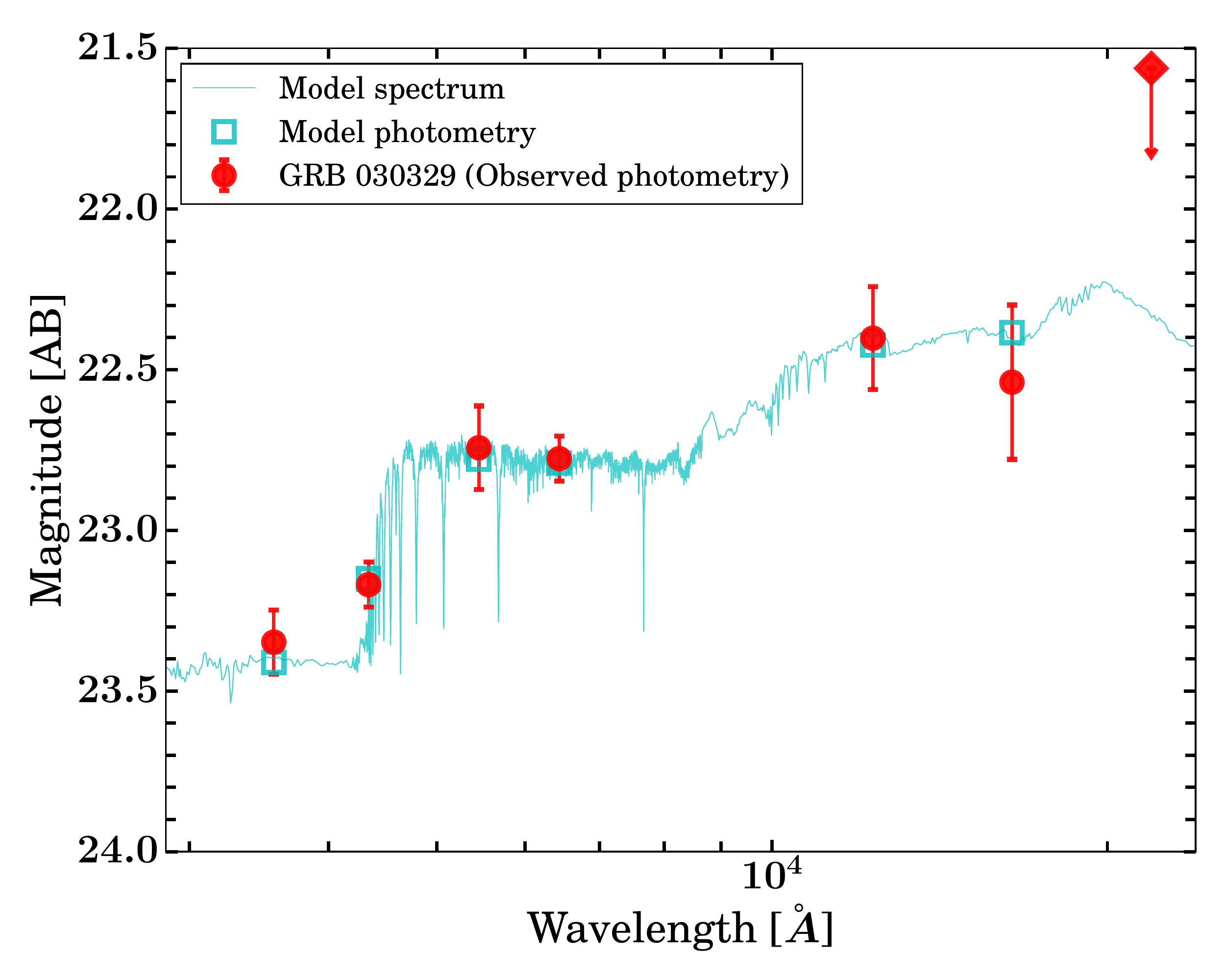}
\includegraphics[width=.9\columnwidth]{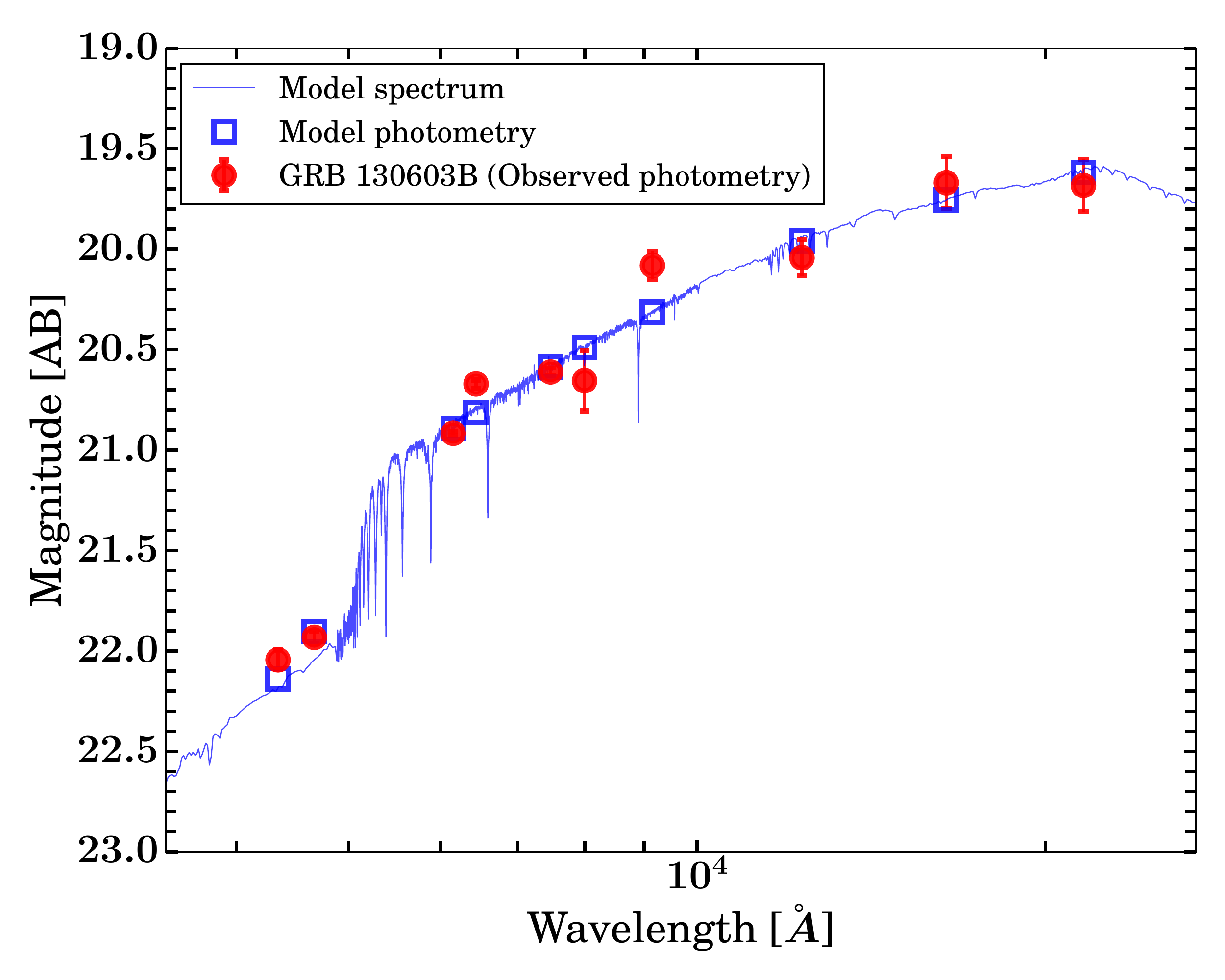}
\includegraphics[width=.9\columnwidth]{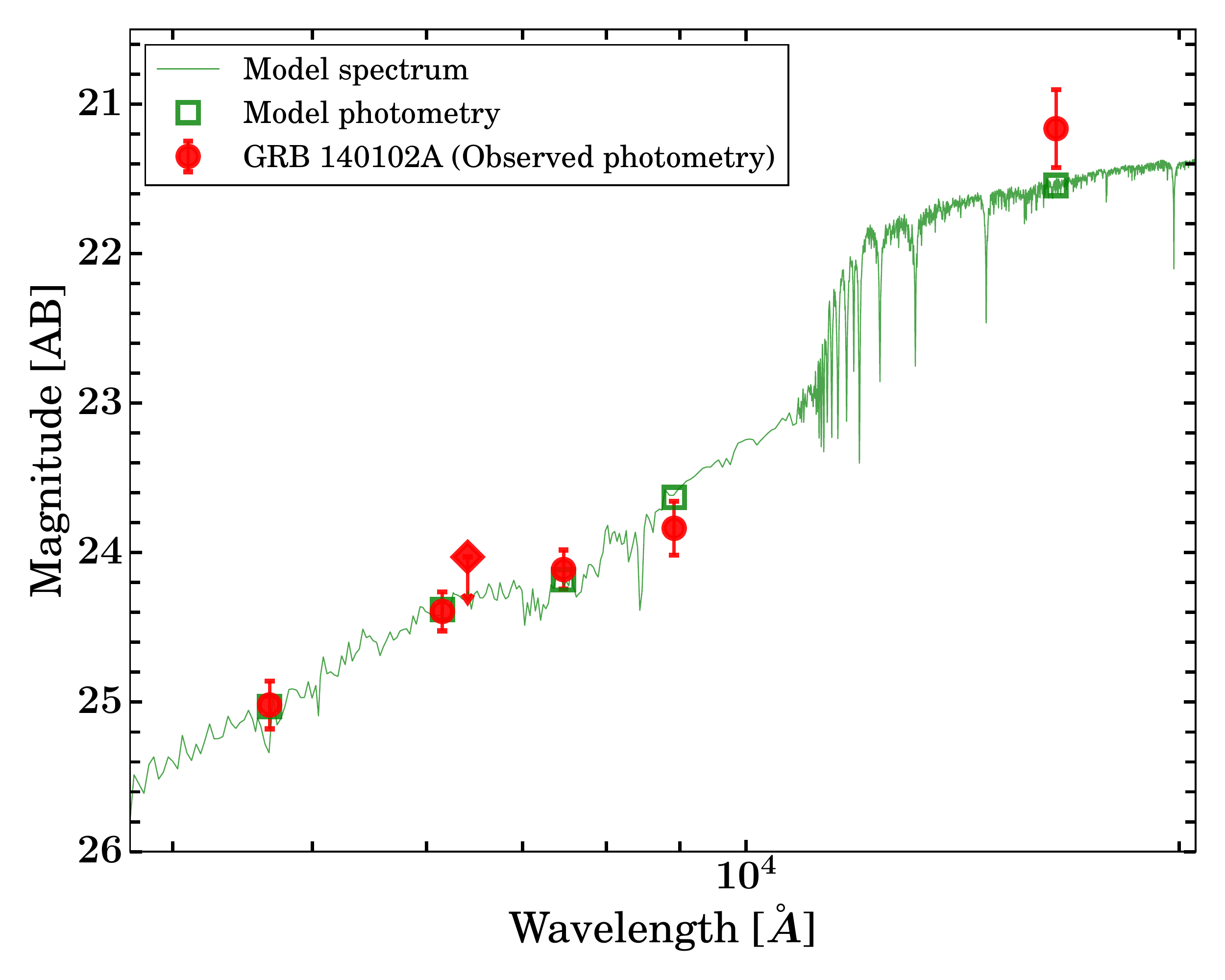}
\includegraphics[width=.9\columnwidth]{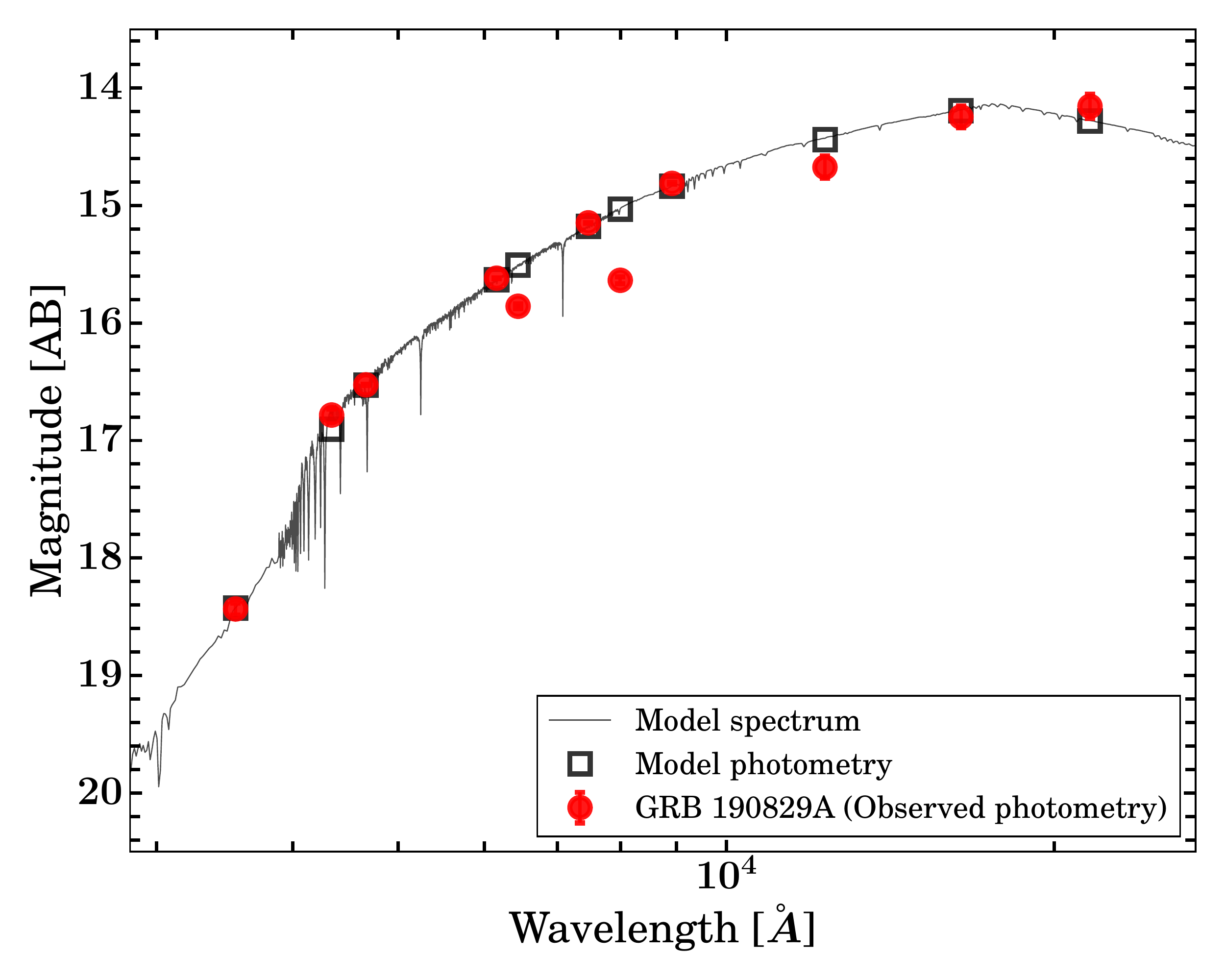}
\includegraphics[width=.9\columnwidth]{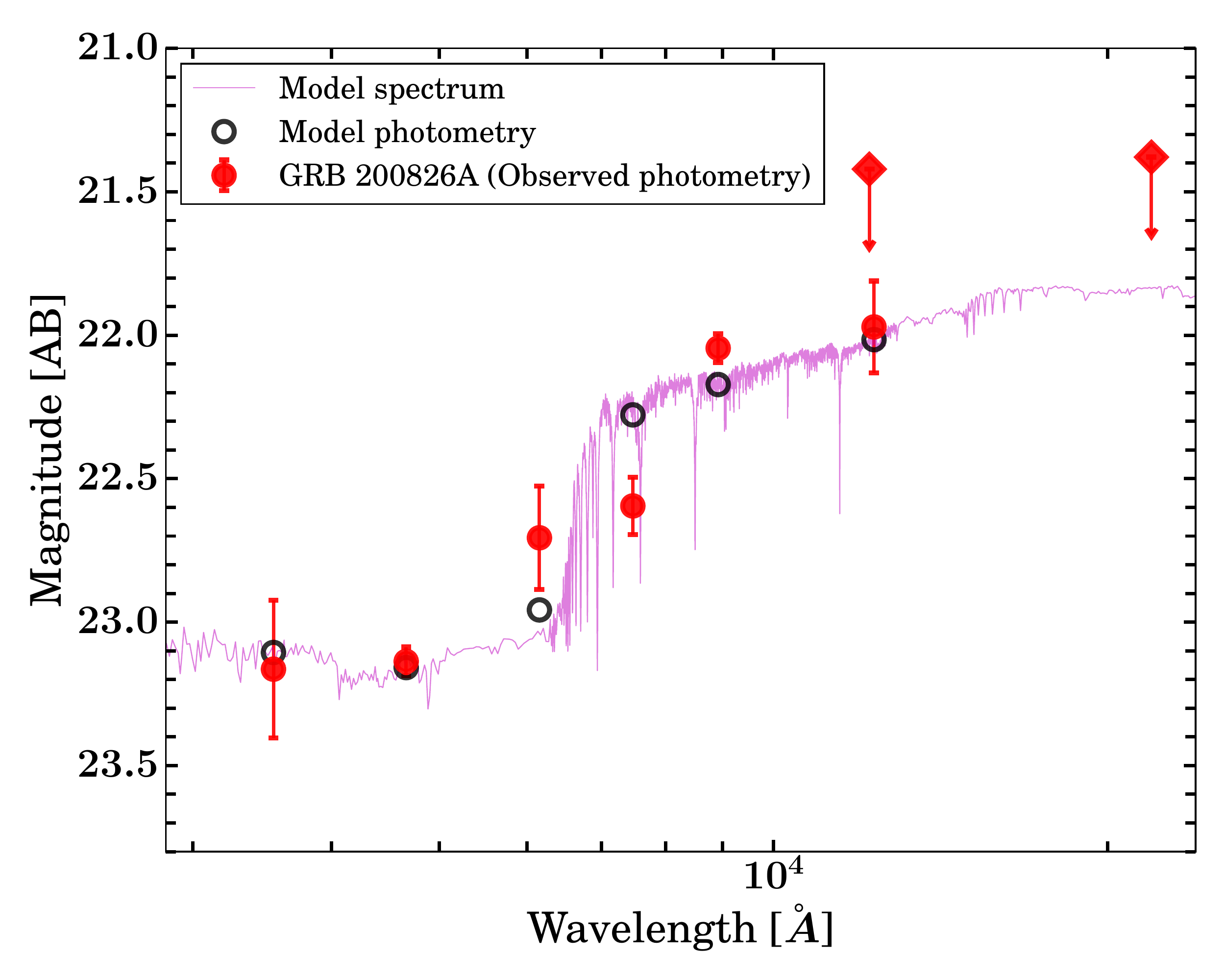}
\caption{{\bf Spectral energy distribution modeling of the host galaxies of our sample:} The best fit modeled spectrum and photometry obtained using \sw{Prospector} are presented in cyan, blue, green, black, magenta for GRB 030329, GRB 130603B, GRB 140102A, GRB 190829A, and GRB 200826A, respectively. The photometry data of the host galaxies (corrected for galactic extinction) in the AB system is shown with red circles.}
\label{sed}
\end{figure*}

\subsubsection{\bf GRB 030329:}

We modeled the SED (constructed using the data taken using 3.6m DOT along with those published in \cite{2005AA...444..711G}) of the host galaxy of GRB 030329 (a low redshift galaxy) using \sw{Prospector}. The best fit SED using nested sampling via \sw{dynesty} method provide the following physical parameters: stellar mass formed (log(M/M$_{\odot}$)) = 7.98$^{+0.12}_{-0.14}$, stellar metallicity ($log (Z/Z_\odot$) = -0.29$^{+0.26}_{-0.18}$, age of the galaxy ($t_{gal}$) = 1.21$^{+0.59}_{-0.44}$ Gyr, rest-frame dust attenuation ($A_{V}$) = 0.12$^{+0.15}_{-0.08}$ mag, and log($\tau$) = 2.11$^{+1.23}_{-1.19}$. Furthermore, we derive the star formation rate of the galaxy = 1.57 $\times 10^{-1}$ M$_{\odot} \rm ~yr^{-1}$ and it is consistent with the value report by \cite{2003Natur.423..847H} based on the host galaxy emission line properties. Our analysis suggests for a low mass, and low star formation galaxy for GRB 030329 \citep{2005AA...444..711G}. 

\subsubsection{\bf GRB 130603B:}

We performed the modeling of the photometric data of the host galaxy obtained using 3.6m DOT along with those taken with other facilities (OSN, CAHA, and GTC) using the \sw{LePHARE} software and presented the results in \citep{2019MNRAS.485.5294P}. We find that the burst's environment is undergoing moderate star formation activity \citep{2019MNRAS.485.5294P}. However, we noted that \sw{LePHARE} software has some major limitations for such analysis, which we discussed in section \ref{sedmodeling}. Therefore, in this work, we performed the SED modeling of GRB 130603B using an advanced software called \sw{Prospector}, and we derive following host parameters: log(M/M$_{\odot}$) = 10.63$^{+0.09}_{-0.10}$, $log(Z/Z_\odot)$ = -1.50$^{+0.40}_{-0.36}$, age of the galaxy = 7.09$^{+1.76}_{-2.14}$ Gyr, $A_{V}$ = 1.65$^{+0.29}_{-0.23}$ mag, and log($\tau$) = 1.78$^{+0.76}_{-0.72}$. We calculated the star formation rate of the galaxy = 11.57 M$_{\odot} \rm ~yr^{-1}$. Our results suggest that the host galaxy of GRB 130603B has a high mass galaxy with a moderate star formation activity, consistent with those reported in \cite{2019MNRAS.485.5294P}.

\subsubsection{\bf GRB 140102A:}

In our recent work \citep{2021MNRAS.505.4086G}, we performed the SED modeling of the host galaxy of GRB 140102A using \sw{LePHARE} software (uses $\chi^{2}$ statistics) with PEGASE2 stellar synthesis population models library. We obtained the best fit solution with the following host galaxy parameters: age of the stellar population in the host galaxy = $9.1 \pm 0.1$ Gyr, mass = ($1.9 \pm 0.2) \times 10^{11}$ M$_{\odot}$, and SFR = $20 \pm 10$ M$_{\odot} \rm ~yr^{-1}$ with a relative poor chi-square value ($\chi^{2}$ = 0.1). This indicates that the error bars are overestimated or the model is too flexible that may cause a large degeneracy in model parameters. For the present work, we collected the photometric observations for the host galaxy of GRB 140102A from our recent work \cite{2021MNRAS.505.4086G}, and performed the modeling using \sw{Prospector} due to the limitation of \sw{LePHARE} software as mentioned above. We have frozen the redshift $z$ = 2.02 obtained from afterglow SED of GRB 140102A, to model the host SED using \sw{Prospector}. We find stellar mass of log(M/M$_{\odot}$) = 11.88$^{+0.34}_{-0.32}$, stellar metallicity of $log (Z/Z_\odot)$ = -0.19$^{+0.78}_{-1.03}$, age of the galaxy ($t_{gal}$) = 8.51$^{+3.30}_{-3.58}$ Gyr, dust extinction of $A_{V}$ = 1.35$^{+0.25}_{-0.25}$ mag, and with a moderate star formation rate of 52.90 M$_{\odot} \rm ~yr^{-1}$. The results indicate that the host was a high-mass galaxy with high star-formation rate, consistent with those obtained from \sw{LePHARE}.

\subsubsection{\bf GRB 190829A:}

GRB 190829A has a very bright and nearby SDSS host galaxy. We modeled the SED using the data observed using 3.6m DOT along with those reported in literature (see Table 2 in appendix). We find stellar mass of log(M/M$_{\odot}$) = 12.04$^{+0.09}_{-0.10}$, stellar metallicity of $log (Z/Z_\odot)$ = -2.39$^{+0.24}_{-0.21}$, age of the galaxy ($t_{gal}$) = 9.91$^{+1.85}_{-2.21}$ Gyr, dust extinction of $A_{V}$ = 2.37$^{+0.22}_{-0.20}$ mag, and with a moderate star formation rate of 6.87 M$_{\odot} \rm ~yr^{-1}$. The SED modeling indicates that the host of GRB 190829A is a massive galaxy with a very high star-formation rate.

\subsubsection{\bf GRB 200826A:}

We modeled the host galaxy of GRB 200826A \citep{2022ApJ...932....1R} using the observed magnitude values using TANSPEC mounted on the main axis of 3.6m DOT along with data published in \cite{2021NatAs.tmp..142A}, and find stellar mass of log(M/M$_{\odot}$) = 9.92$^{+0.08}_{-0.10}$, stellar metallicity of $log(Z/Z_\odot)$ = -0.37$^{+0.19}_{-0.21}$, age of the galaxy ($t_{gal}$) = 4.74$^{+1.53}_{-1.90}$ Gyr, dust extinction of $A_{V}$ = 0.19$^{+0.17}_{-0.11}$ mag, and with a moderate star formation rate of 3.49 M$_{\odot} \rm ~yr^{-1}$. These parameters are typical to those observed for long GRBs host galaxies and consistent with \cite{2021NatAs.tmp..142A}.

\subsection{\bf Comparison with known sample of host galaxies}

\begin{figure*}[!t]
\includegraphics[scale=0.35]{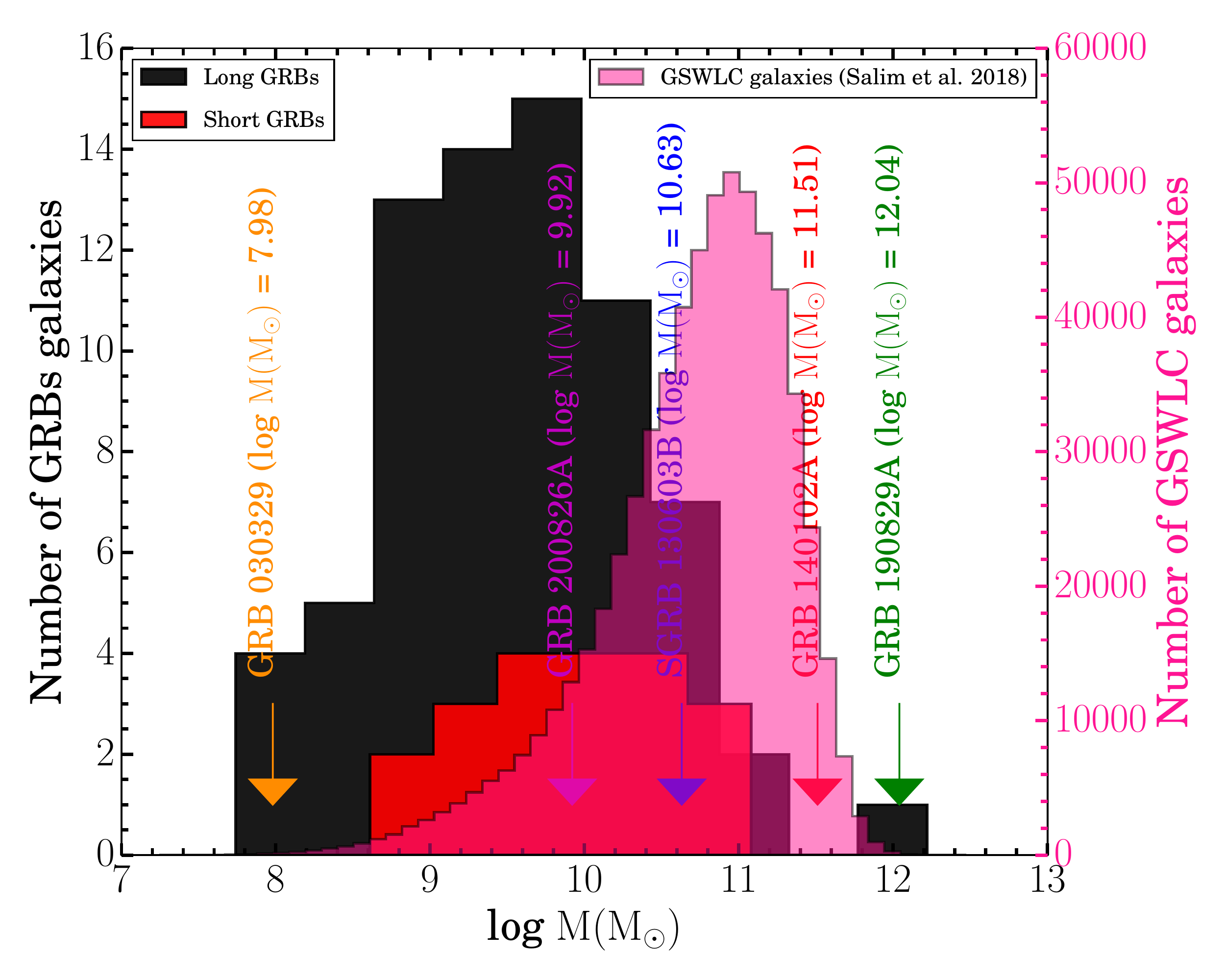}
\includegraphics[scale=0.35]{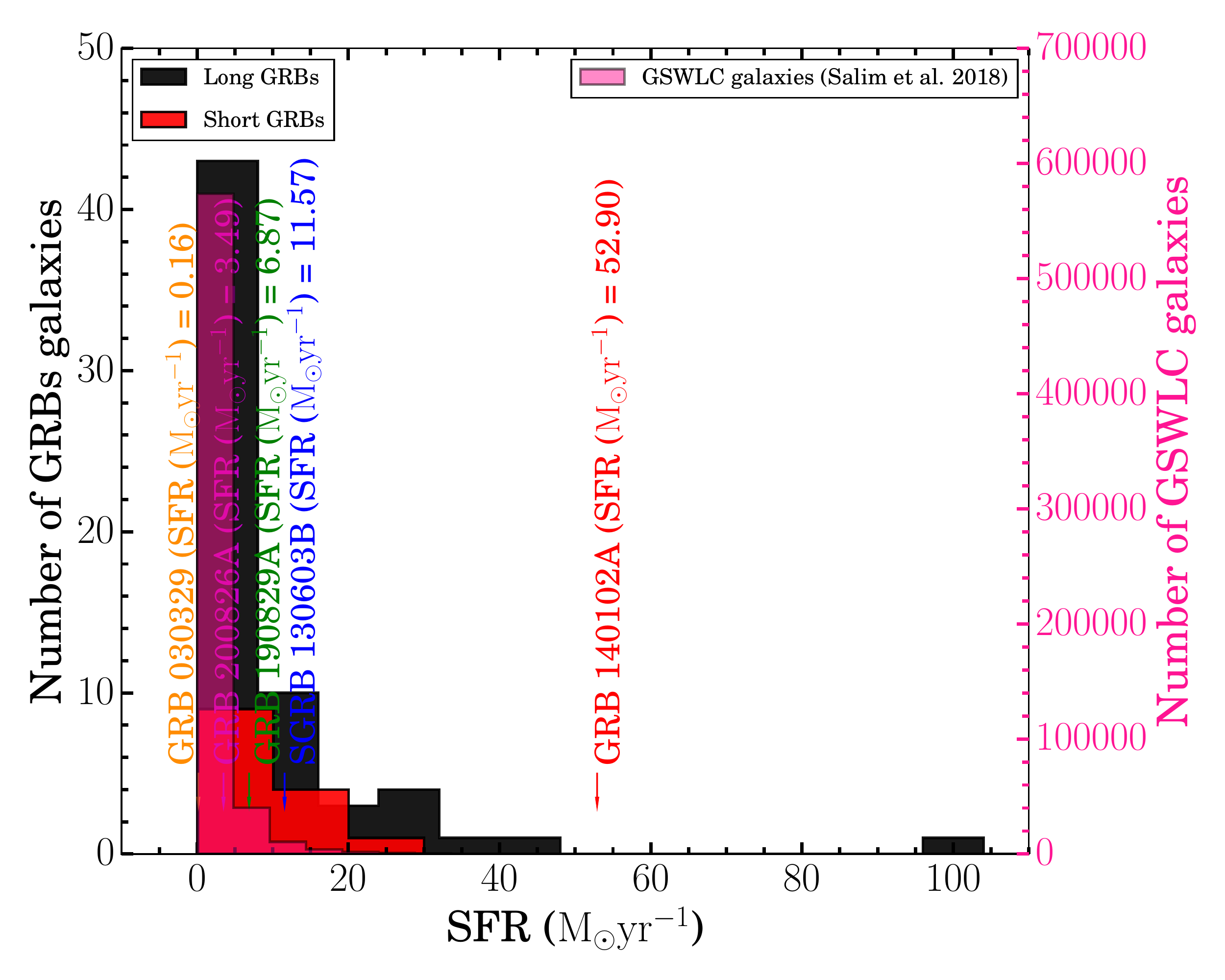}
\includegraphics[scale=0.26]{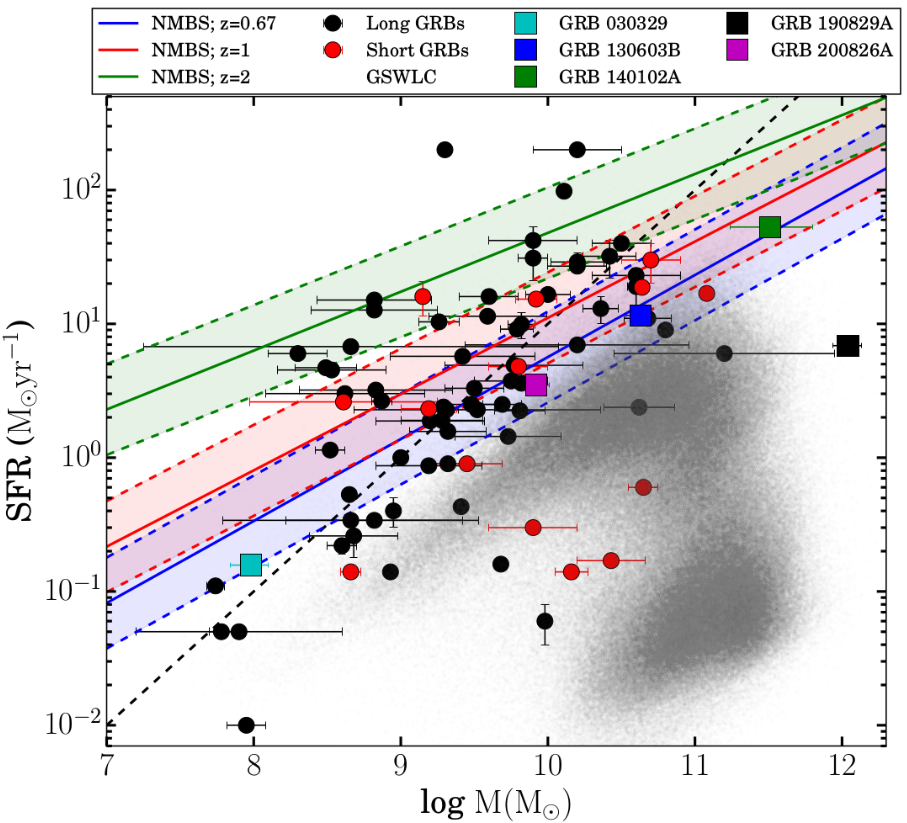}
\includegraphics[scale=0.26]{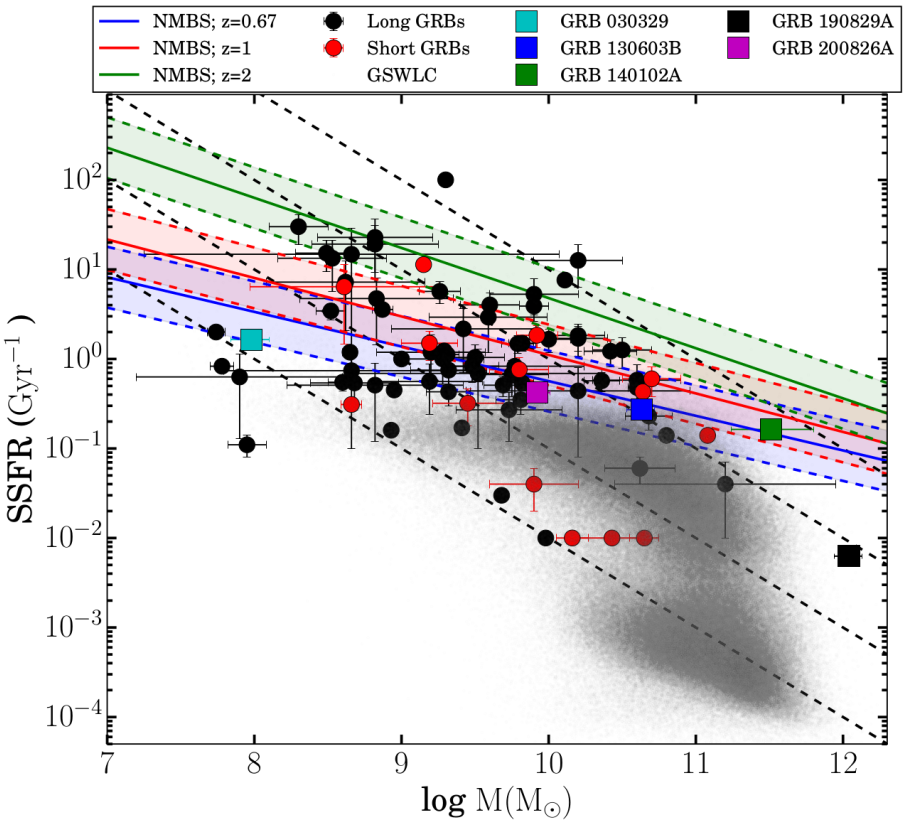}
\caption{{\it Top left panel:} The distribution of stellar mass for the host galaxies of GRBs and galaxies studied in GALEX-SDSS-WISE Legacy Catalog (right side Y-scale). {\it Top right panel:} The distribution of SFR for the host galaxies of GRBs and galaxies studied in the GALEX-SDSS-WISE Legacy Catalog (right side Y-scale). The positions of the host galaxies presented in our sample are shown with downwards arrows. {\it Bottom left panel:} The stellar mass of the host galaxies versus star formation rate for our sample, obtained from the host SED modeling. Square markers indicate the host galaxies' position for our sample. The dashed line indicates a constant specific star formation rate of 1 Gyr$^{-1}$. {\it Bottom right panel:} The stellar mass of the host galaxies versus specific star formation rate for our sample. Square markers indicate the host position for our sample. The dashed lines indicate the constant star formation rates of 0.1, 1, 10, and 100 M yr$^{-1}$ from left to right. Black and red circles show the long GRBs and short GRBs host galaxies with star formation rates (bottom left) and specific star formation rates (bottom right) calculated from GHostS during 1997 to 2014 \citep{2006AIPC..836..540S, 2009ApJ...691..182S}. The grey dots show the location of GSWLC galaxies in both the correlations \citealt{2016ApJS..227....2S, 2018ApJ...859...11S}. The colored solid lines and shaded regions in both the correlations indicate the best fit power-law functions and their dispersion calculated for normal star-forming galaxies from NEWFIRM Medium-Band Survey (NMBS) sample at redshift values of 0.67 (the mean redshift of the sample), 1 and 2 (the mean redshift of long GRBs), respectively \citep{2012ApJ...754L..29W}.}
\label{host_sfr}
\end{figure*}

We compared the host galaxy's properties of our sample (see Table \ref{tab:sedresults}) with other well-studied samples of GRB host galaxies from the literature \citep{2006AIPC..836..540S, 2009ApJ...691..182S} and GALEX-SDSS-WISE Legacy Catalog\footnote{This catalog comprises the properties of $\sim$ 700,000 galaxies with measured redshifts values below 0.3 using SDSS.} (GSWLC; \citealt{2016ApJS..227....2S, 2018ApJ...859...11S}). The physical parameters (stellar mass and star formation rate) distribution for the host galaxies of GRBs and GSWLC are shown in Fig. \ref{host_sfr} (upper left and right panels). The positions of the host galaxies presented in our sample are shown with downwards arrows. We noticed that the average value of the stellar mass of the galaxies in the GSWLC is higher than the average values of the stellar mass of the host galaxies of long and short GRBs. On the other hand, the average value of star formation rate (SFR) of the galaxies in the GSWLC is lower than the average values of the stellar mass of the host galaxies of long and short GRBs. Furthermore, we studied possible correlations of the stellar mass of the host galaxies versus SFR and the stellar mass of the host galaxies versus specific star formation rates (SSFRs). For normal star-forming galaxies, the correlation between the stellar mass and SFR is defined as the "main sequence." This correlation reveals the possible procedures of the star formation histories of galaxies. If the correlation is tighter, it suggests that the star formation history traces stellar mass growth more smoothly. On the other hand, if the correlation is weaker (high scatter), it suggests a random star formation history \citep{2007ApJ...670..156D, 2007ApJ...660L..43N, 2011MNRAS.410.1703F}. Hence, GRB host galaxy properties can be characterized by comparing them with the main sequence. The relation between the stellar mass of the host galaxies as a function of SFRs for GRBs and GSWLC is shown in Fig. \ref{host_sfr} (bottom left panel). The position of the host galaxies presented in our sample is shown with colored squares. The colored solid lines and shaded regions in both the correlations indicate the best fit power-law functions and their dispersion calculated for normal star-forming galaxies from NEWFIRM Medium-Band Survey (NMBS) sample at redshift values of 0.67 (the mean redshift of the sample), 1 and 2 (the mean redshift of long GRBs), respectively \citep{2012ApJ...754L..29W}. We noticed that the host galaxies' physical properties of GRBs are more common to normal star-forming galaxies at the high-redshift Universe in comparison to the low-redshift Universe \citep{2013ApJ...778..128P, 2019MNRAS.488.5029H}.

In addition, we found that GRBs in our sample follow Mass-SFR correlation (see Fig. \ref{host_sfr}) as described previously by \cite{2006AIPC..836..540S, 2009ApJ...691..182S}. Furthermore, we notice that the star formation rate of GRB 130603B, GRB 140102A, GRB 190829A, and GRB 200826A in our sample are higher in comparison to the median value of 2.5 M$_{\odot} \rm ~yr^{-1}$ \citep{2009ApJ...691..182S}. On the other hand, GRB 030329 has a low star formation rate. The host galaxies of GRB 130603B, GRB 140102A, GRB 190829A, and GRB 200826A have higher mass than the galaxies with semi-star formation rates. We also compared the SSFRs of the host galaxies of our sample with the GRB's host galaxies sample studied by \cite{2006AIPC..836..540S, 2009ApJ...691..182S} and normal star-forming galaxies (GSWLC; \citealt{2016ApJS..227....2S, 2018ApJ...859...11S}). The SSFR indicates the intensity of star formation in particular galaxies. The correlation between the stellar mass of the galaxies and SSFRs suggests how the galaxies compose their stellar populations \citep{2015A&A...577A.112L}. We noticed that other than GRB 030329, all other four host galaxies have lower SSFR in comparison to the average value of 0.8 G$\rm yr^{-1}$ (see Fig. \ref{host_sfr}), suggesting a lower intensity of star-formation for these host galaxies. On the other hand, the observed higher value specific star formation rate for the host galaxy of GRB 030329 indicates a young, starbursting galaxy \citep{2006ApJ...653L..85C}. The relation between SSFR-Mass also indicates that physical properties of the host galaxies of GRBs are more common to normal star-forming galaxies at the high-redshift \citep{2013ApJ...778..128P, 2019MNRAS.488.5029H}.

\subsubsection{\bf Host galaxies of GRBs, and Supernovae:}

\begin{figure}[!t]
\includegraphics[scale=0.35]{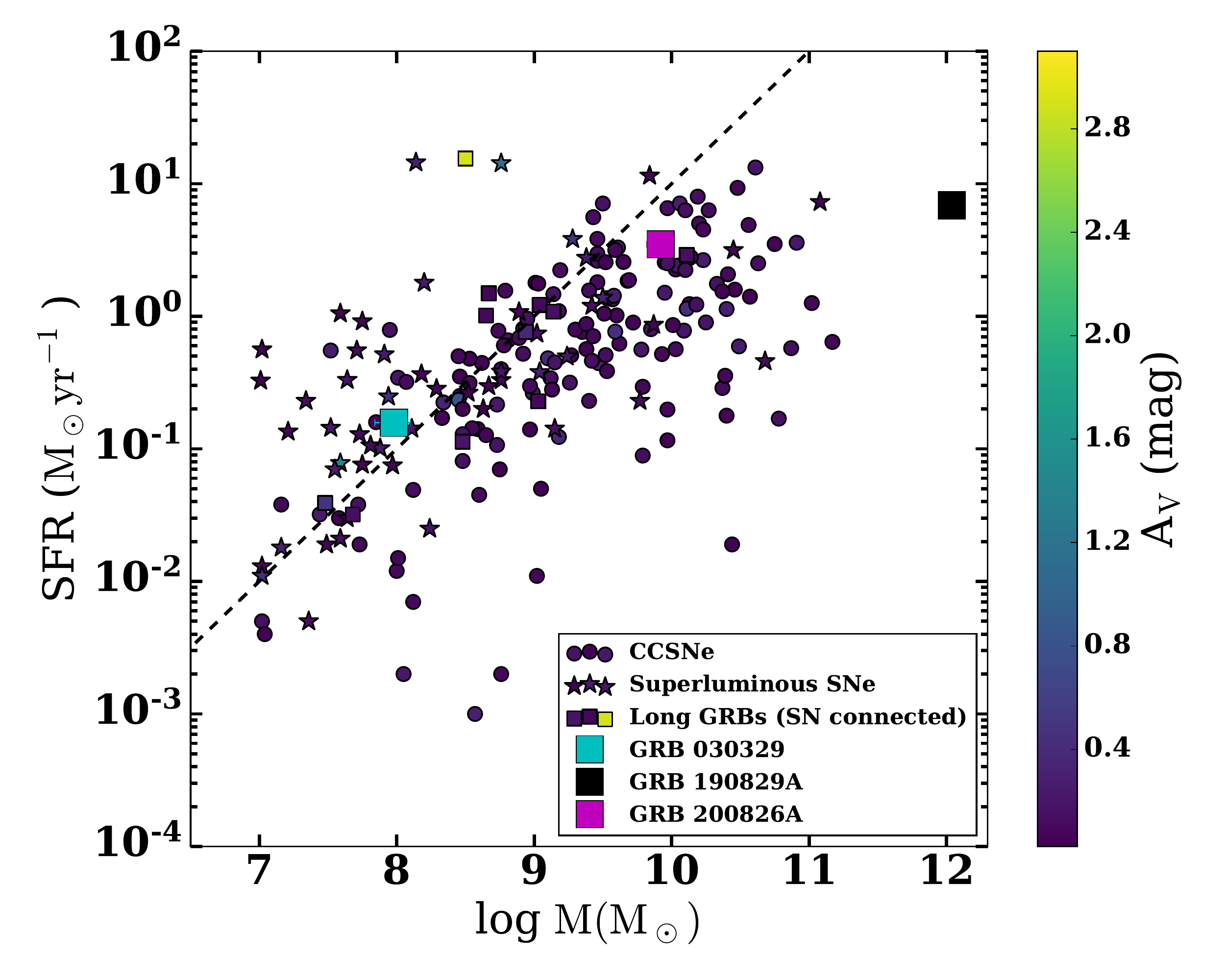}
\caption{The stellar mass of the host galaxies versus star formation rate for GRBs connected with supernovae in our sample, obtained from the host SED modeling. Circles, stars, and squares markers indicate the position of host galaxies of CCSN, superluminous supernovae, and long GRBs connected with supernovae, respectively, taken from \cite{2021MNRAS.503.3931T}. The right side Y-scale shows the corresponding A$_{V}$ values.  
}\label{host_sfr_2021}
\end{figure}

Long GRBs usually occur at high redshift; however, some of the nearby long bursts are associated with broad-line type Ic supernovae (stripped-envelope). However, it is still not understood that all long GRBs are connected with broad-line type Ic supernovae, and we could only detect the near ones due to the observational constraints \citep{2017AdAst2017E...5C}. Therefore, the examination of the host galaxies properties of long GRBs and supernovae will be helpful to explore their environment and progenitors. Recently, \cite{2021MNRAS.503.3931T} compared the host galaxy properties of long bursts with core-collapse supernovae and superluminous supernovae and suggested that cumulative properties of the host galaxies of long GRBs without supernovae and with supernovae are not much different. Out of four long GRBs in our sample, three GRBs (GRB 030329/SN 2003dh, GRB 190829A/ SN 2019oyw, and GRB 200826A) were associated with broad-line type Ic supernovae. We compared (mass as a function of SFR) the results of these GRBs/SNe with those published in \cite{2021MNRAS.503.3931T}. We find that GRB 030329 and GRB 200826A follow the correlation plane of long GRBs and CCSN; however, the host galaxy of GRB 190829A lies on the right side of the distribution (see Fig. \ref{host_sfr_2021}). We searched the host galaxy (SDSS) of GRB 190829A in GSWLC sample studied by \cite{2016ApJS..227....2S, 2018ApJ...859...11S} and found that the SFR and Stellar-mass values of the host galaxy (ObjID: 1237652899156721762) of GRB 190829A are log (SFR)= 0.395 $\pm$ 0.103 M$_{\odot} \rm ~yr^{-1}$, log(M/M$_{\odot}$ = 11.256 $\pm$ 0.012, respectively {\footnote{The model fit used by \cite{2016ApJS..227....2S, 2018ApJ...859...11S} is underfitting the observed data (reduced chi-square = 2.15)}}. These values also indicate that the host galaxy of GRB 190829A is a massive and high star-forming galaxy, consistent with our results.

\subsection{\bf Dust and gas in the host galaxies:}

\begin{figure}[!ht]
\centering
\includegraphics[width=.88\columnwidth]{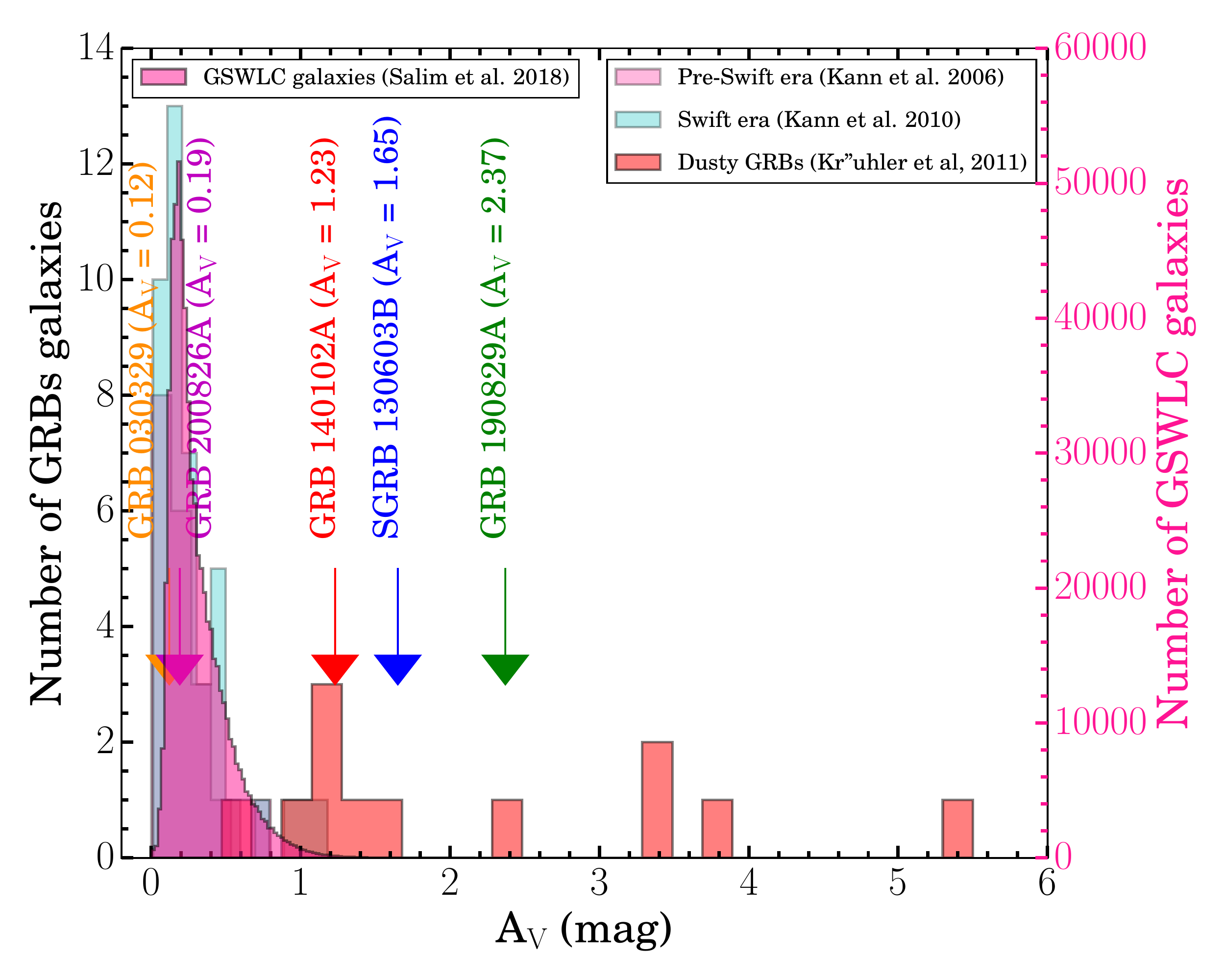}
\includegraphics[width=.88\columnwidth]{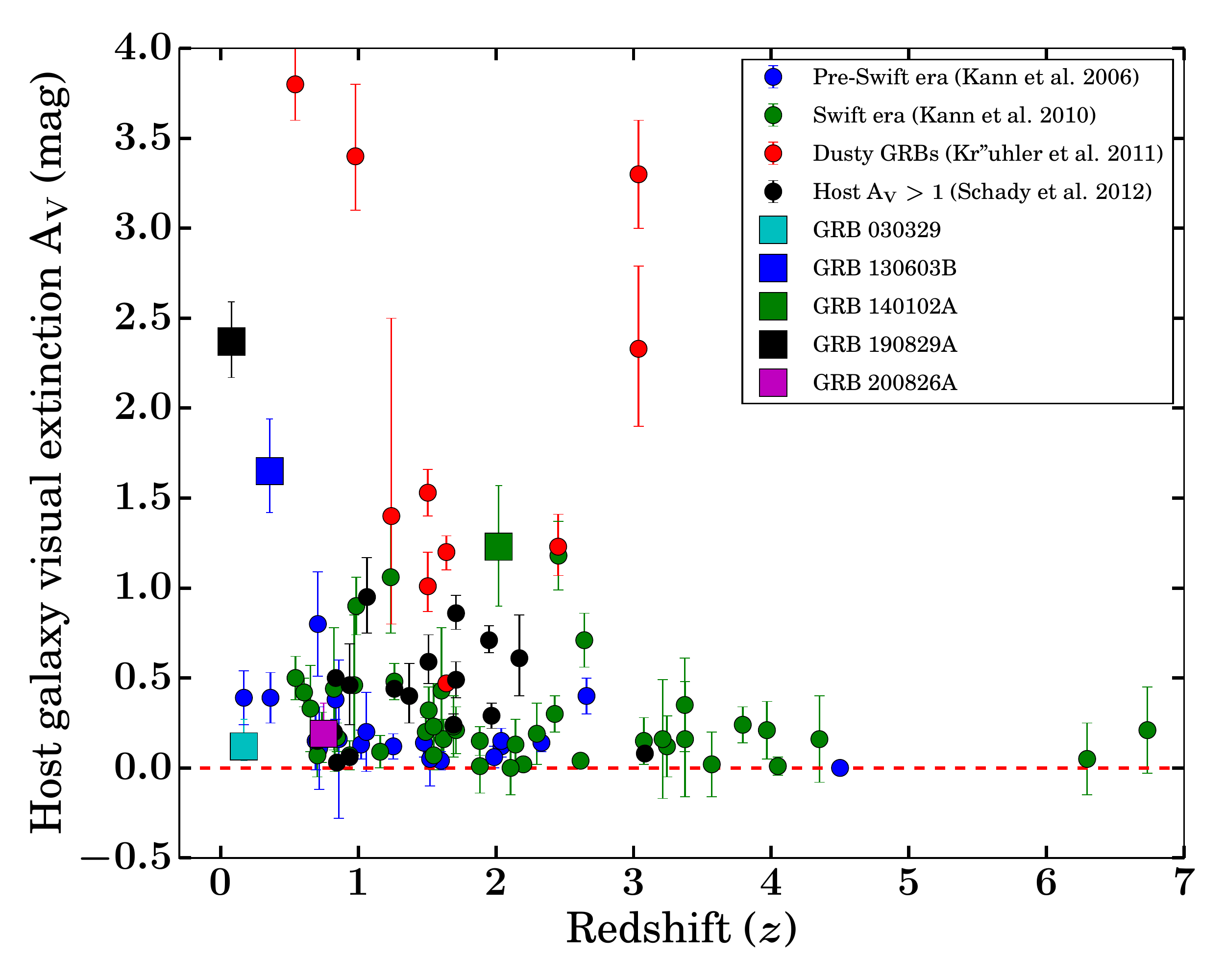}
\includegraphics[width=.88\columnwidth]{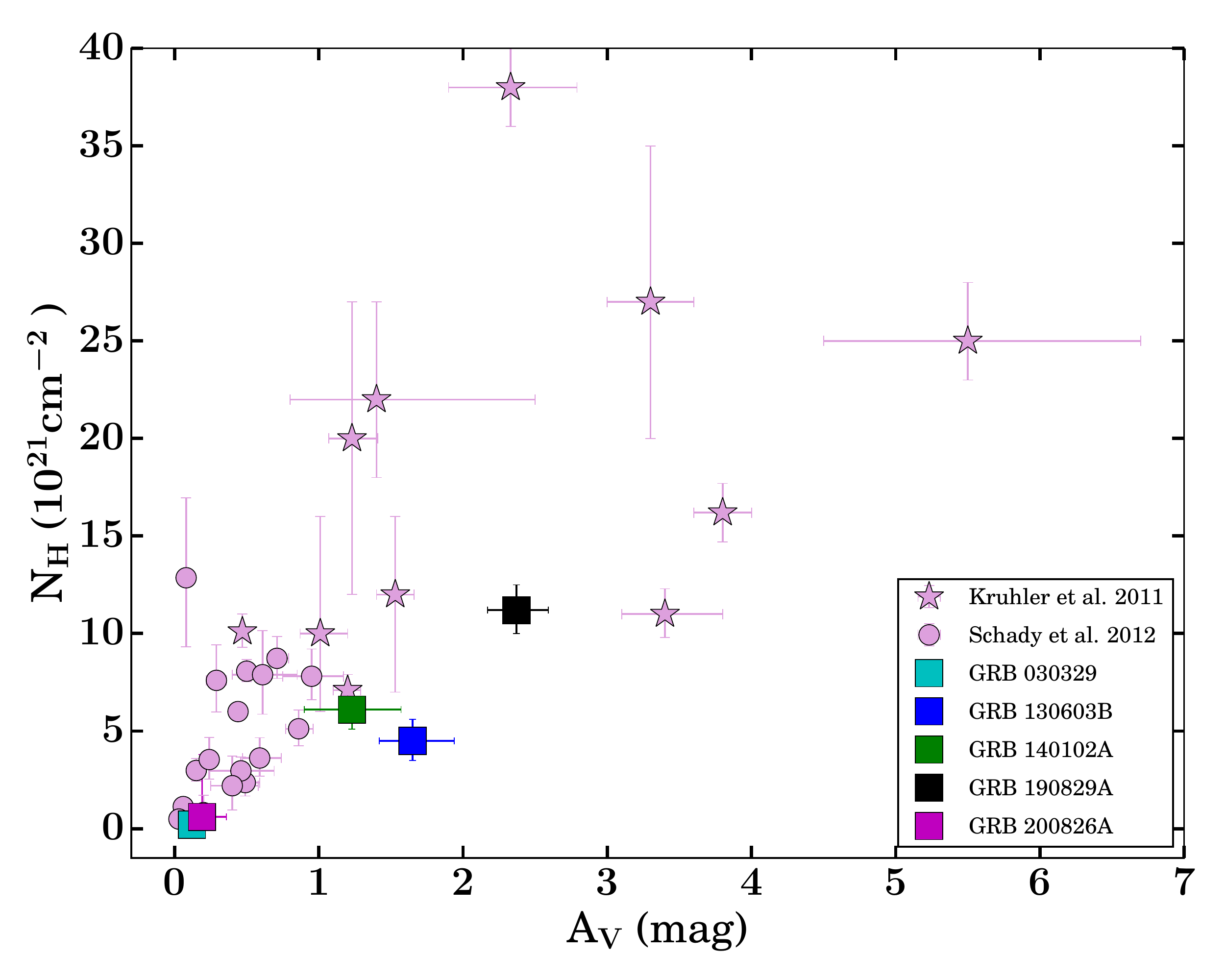}
\caption{{Top panel:} The distribution of visual extinction (in the source frame) of the host galaxies in our sample. For the comparison, the data points for GRBs in pre-\swift \citep{2006ApJ...641..993K}, post-\swift \citep{2010ApJ...720.1513K} era, dusty GRBs \citep{2011A&A...534A.108K} and GSWLC \citep{2016ApJS..227....2S, 2018ApJ...859...11S} are also shown. {Middle panel:} Redshift evolution of visual extinction. The horizontal red dashed line shows A$_{v}$ = 0. {Bottom panel:} The dust extinction as a function of X-ray column density in the local environment of the host galaxies of our sample, along with other data points taken from \cite{2011A&A...534A.108K, 2012A&A...537A..15S}.}
\label{host_av_nh}
\end{figure}

We calculated the dust extinction in the local environment of the host galaxies of our sample and compared them with a larger sample of host galaxy of GRBs \citep{2006ApJ...641..993K, 2010ApJ...720.1513K, 2011A&A...534A.108K}/ GSWLC \citep{2016ApJS..227....2S, 2018ApJ...859...11S} galaxies taken from the literature. We determine the visual dust extinction in rest-frame (A$_{v}$) using the host galaxy SED modeling of each burst in our sample (see section \ref{sedmodeling} for more details). Fig. \ref{host_av_nh} (top panel) shows the distribution of visual extinction (in the source frame) of the host galaxies in our sample. For the comparison, we have also shown the distribution of visual extinction for GRBs in pre-\swift \citep{2006ApJ...641..993K}, post-\swift \citep{2010ApJ...720.1513K} era, dusty GRBs \citep{2011A&A...534A.108K} and GSWLC \citep{2016ApJS..227....2S, 2018ApJ...859...11S}. We find that A$_{v}$ values are distributed over a wide range for our sample, and part of host galaxies (GRB 130603B, GRB 140102A, and GRB 190829A) are extinguished by dust. Moreover, these galaxies (dusty) typically have higher stellar mass, consistent with previous studies \citep{2009AJ....138.1690P, 2011A&A...534A.108K}. In addition, we also noted that the A$_{v}$ values $\leq$ 1 for other than the dusty sample, suggesting that the dusty galaxies ($\sim$ 20 \%) are highly extinguished and might cause optical darkness \citep{2004ApJ...617L..21J, 2011A&A...526A..30G, 2021arXiv211111795G}.

Fig. \ref{host_av_nh} (middle panel) shows the evolution of host visual extinction as a function of redshift for our sample along with those data points published by \citep{2006ApJ...641..993K, 2010ApJ...720.1513K, 2011A&A...534A.108K, 2012A&A...537A..15S}. We noticed that the visual extinction is decreasing with redshift, although it might be due to the selection effect as there are only a few GRBs with redshift $\geq$ 5, and the observation of dusty galaxies at such high redshift is very difficult \citep{2010ApJ...720.1513K}.   

Furthermore, we used the spectral analysis results of the X-ray afterglow data from the literature to constrain the intrinsic hydrogen column density (N$_{\rm H}$ ($z$)) of each burst in our sample. The X-ray afterglow spectra of GRBs could be typically described using a simple absorption power-law model consists of three components: a Galactic absorption (N$_{\rm H}$), host absorption (N$_{\rm H}$ ($z$)), and a power-law component due to synchrotron emission. We obtained N$_{\rm H}$ ($z$) values of GRB 030329 from \cite{2004A&A...423..861T}, GRB 130603B from \swift X-ray telescope web-page maintained by Phil Evans, GRB 140102A from \cite{2021MNRAS.505.4086G}, GRB 190829A from \cite{2020ApJ...898...42C}, and for GRB 200826A from \cite{2021NatAs.tmp..142A}, respectively. The distribution of host dust extinction and gas column densities in the local environment for our sample, along with other data points taken from the literature, are shown in Fig. \ref{host_av_nh} (bottom panel). We noticed that GRB 190829A has a considerable amount of dust and gas in the local environment of its host galaxy. The observed considerable amount of dust and gas might be related to the associated VHE emission from GRB 190829A, a similar dusty environment is also seen in the case of other VHE detected bursts such as GRB 190114C \citep{2020A&A...633A..68D} and GRB 201216C \citep{2022MNRAS.tmp.1037R}. However, due to the limited number of VHE detected GRBs, it is still an unsolved problem whether VHE detected GRBs require unique environments to emit the VHE emission or VHE emission is only due to the burst emission mechanisms such as Synchrotron Self Compton \citep{2019Natur.575..455M, 2019Natur.575..464A, 2020A&A...633A..68D, 2021RMxAC..53..113G}.

\begin{table*}
\scriptsize
\caption{The stellar population properties of the host galaxies of our sample were obtained using spectral energy density modeling using \sw{Prospector}. NH$_{\rm host}$ and log(e) denote the intrinsic column density and evidence values, respectively.}
\begin{center}
\begin{tabular}{c c c c c c c c c c}
\hline
\bf GRB &\bf RA (J2000) & \bf DEC (J2000) &$\bf  z$ & \bf log(M/M$_{\odot}$) & \bf log(Z/Z$_{\odot}$) & $\bf A_{V}$ & $\bf t_{gal}$ (Gyr) & $\bf  log(e)$ & \bf  NH$_{\rm host}$ ($\rm \bf cm^{-2}$)\\ \hline
030329 & 161.21 & 21.52 & 0.1685 &7.98$^{+0.12}_{-0.14}$ & -0.29$^{+0.26}_{-0.18}$& 0.12$^{+0.15}_{-0.08}$ & 1.21$^{+0.59}_{-0.44}$& 128.95& 2 $\rm \times 10^{20}$\\
130603B &172.20 &17.07 & 0.356&10.63$^{+0.09}_{-0.10}$ & -1.50$^{+0.40}_{-0.36}$& 1.65$^{+0.29}_{-0.23}$ &7.09$^{+1.76}_{-2.14}$ &169.89 & 4.5 $\rm \times 10^{21}$\\
140102A &211.92 &1.33 & 2.02 &11.51$^{+0.29}_{-0.27}$ &-0.27$^{+0.95}_{-1.16}$ &1.23$^{+0.34}_{-0.33}$ & 2.23$^{+0.69}_{-0.81}$&106.06 &6.1 $\rm \times 10^{21}$ \\
190829A & 44.54& -8.96& 0.0785 &12.04$^{+0.09}_{-0.10}$ & -2.39$^{+0.24}_{-0.21}$ &2.37$^{+0.22}_{-0.20}$ &9.91$^{+1.85}_{-2.21}$ & -447.91&1.12 $\rm \times 10^{22}$ \\
200826A &6.78 &34.03 & 0.748&9.92$^{+0.08}_{-0.10}$ &-0.37$^{+0.19}_{-0.21}$ &  0.19$^{+0.17}_{-0.11}$&4.74$^{+1.53}_{-1.90}$ & 118.51&6 $\rm \times 10^{20}$ \\

\hline
\vspace{-2em}
\end{tabular}
\end{center}
\label{tab:sedresults}
\end{table*}

\vspace{-2em}
\section{Summary and Conclusion}
\label{Summary and Conclusion}

The observed gamma-ray prompt emission properties of GRBs do not always depict about nature of their progenitors and environments and, in turn, about unambiguous classification. Recently, the origin of a few short bursts (e.g., GRB 090426 and GRB 200826A) from the collapse of massive stars \citep{2009A&A...507L..45A, 2021NatAs.tmp..142A} and long GRBs (e.g., GRB 211211A, and GRB 060614) from the merger of two compact objects \citep{2022arXiv220903363T, 2022arXiv220410864R, 2015NatCo...6.7323Y} are confirmed. These examples suggest that at least some of the short GRBs might be originated from collapsars, and some of the long GRBs might be originated from compact mergers. Therefore, the late-time observations of the host galaxies are crucial in examining the burst environment and, in turn, the possible progenitors, especially for the hybrid cases. In this paper, we present the photometric observations of the five interesting GRBs' host galaxies observed using India's largest optical telescope (3.6m DOT) to constrain the environment of these bursts, nature of possible progenitors, and explore the deep observations capabilities of 3.6m DOT. Our optical-NIR multi-band data of these five hosts, along with those published ones, were used to perform multi-band modeling of the host galaxies using \sw{Prospector} software (version 1.1.0). We noted that the host galaxies in our sample have a wide range of physical parameters (see Table \ref{tab:sedresults}). The host galaxies of GRB 130603B, GRB 140102A, GRB 190829A, and GRB 200826A have a massive stellar population galaxy with a high star formation rate. However, GRB 030329 has a low-mass galaxy with a low star formation rate, such host galaxies having a low-mass with a low star formation rate are rare \citep{2010ApJ...721.1919C}. We compared the stellar population properties (such as SFR, SSFR, Mass, etc.) of the host galaxies of our sample with a large sample of long and short bursts along with those taken from literature specifically with GSWLC. We find that all the bursts in our sample satisfy the typical known correlation between host galaxy parameters. We noted that the GRBs generally occur in host galaxies that have less massive and high star-forming galaxies than GSWLC galaxies. Further, the host galaxies' physical properties of GRBs are more common to normal star-forming galaxies at the higher redshifts.

In addition, we obtained the X-ray hydrogen column densities from the X-ray afterglow observations of these bursts and studied its distribution with optical dust extinction. We find that GRB 190829A has a considerable amount of dust and gas in the local environment of its host galaxy. A dusty environment is also seen in the case of other VHE detected bursts such as GRB 190114C \citep{2020A&A...633A..68D} and GRB 201216C \citep{2022MNRAS.tmp.1037R}. It suggests that VHE detected GRBs might require a unique local environment for VHE emission to occur.  Unfortunately, due to the small size of the present sample, it is difficult to quantify the selection effects, which further limits a robust statistical analysis. Our results demonstrate that the back-end instruments (such as IMAGER and TANSPEC) of 3.6m DOT have a unique capability for optical-NIR deep observations of faint objects such as host galaxies of GRBs and other interesting transients in the near future. Also, in the near future, systematic studies (with a larger sample) of the host galaxies along with prompt emission and afterglow properties of hybrid GRBs may play a crucial role in understanding their progenitors.


\appendix

\begin{figure*}[!ht]
\centering
\includegraphics[width=.9\columnwidth]{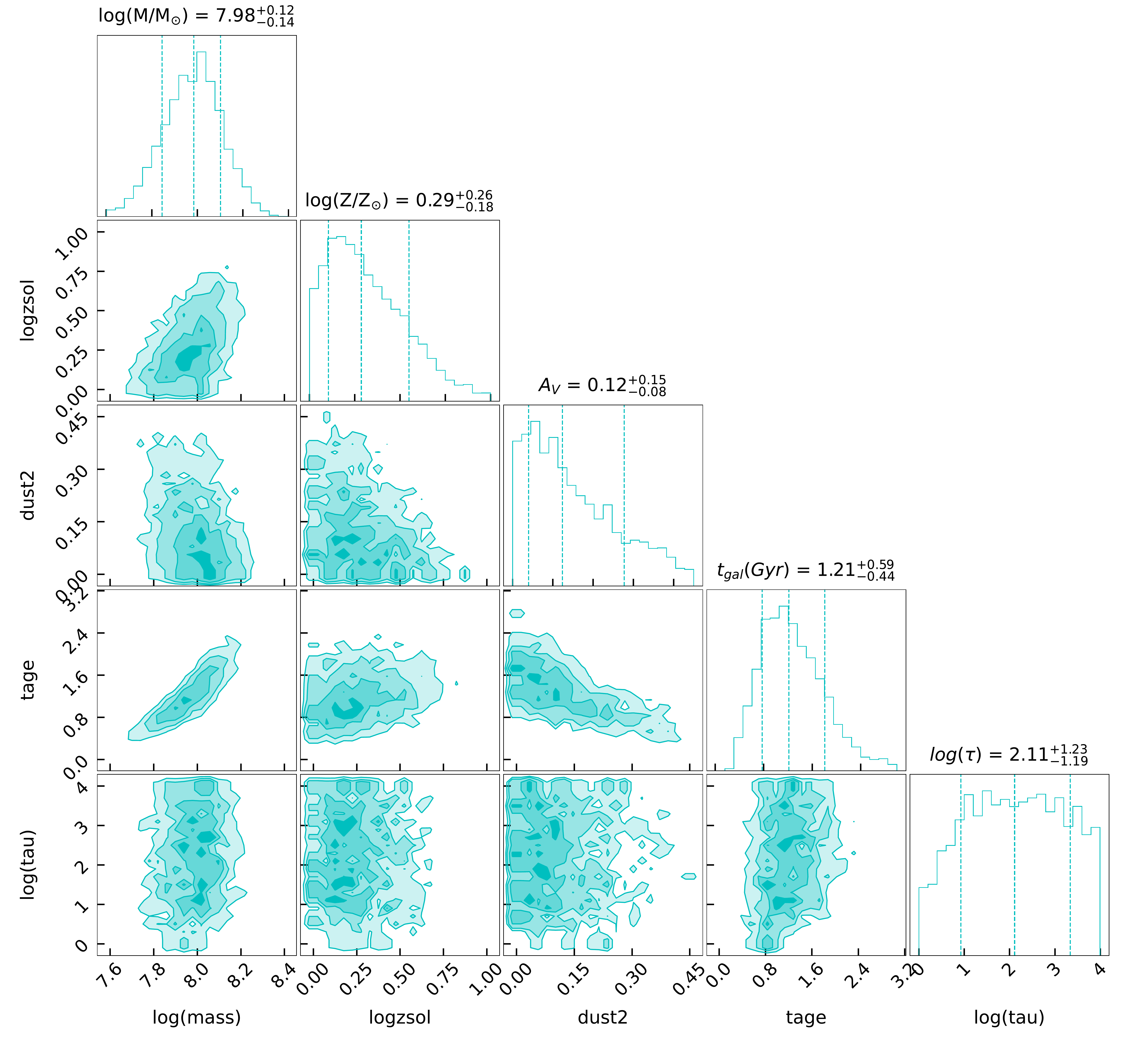}
\includegraphics[width=.9\columnwidth]{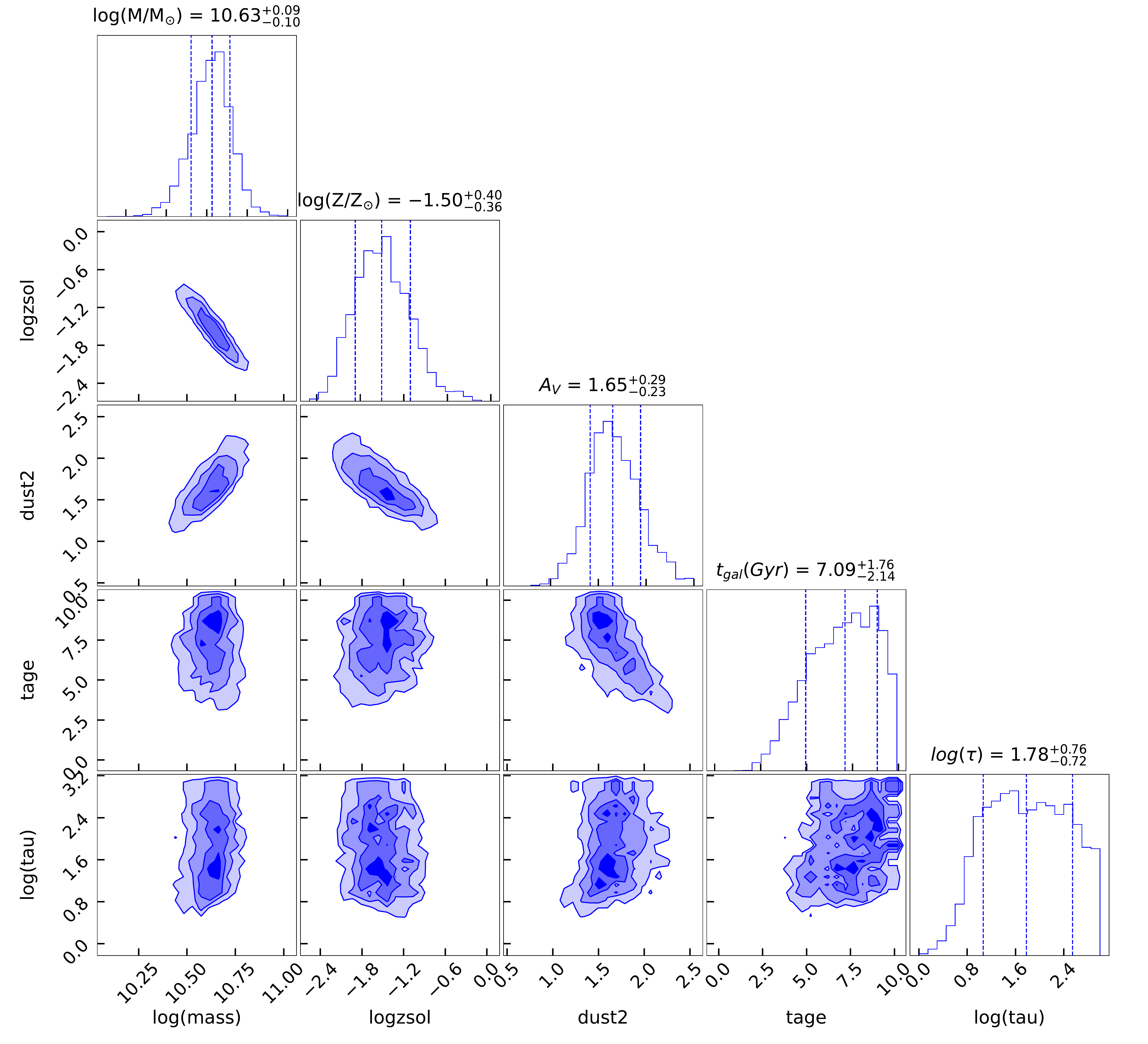}
\includegraphics[width=.9\columnwidth]{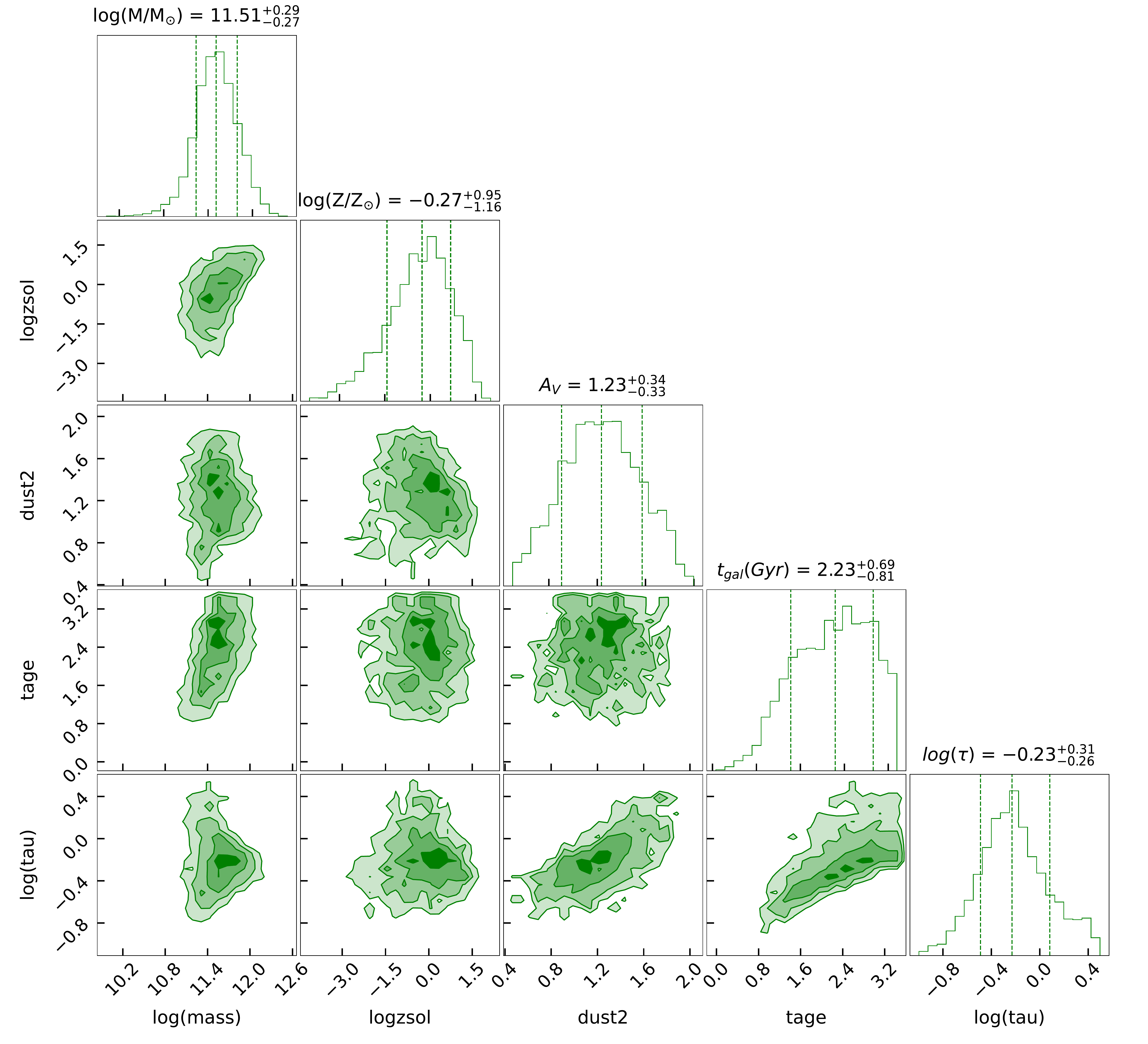}
\includegraphics[width=.9\columnwidth]{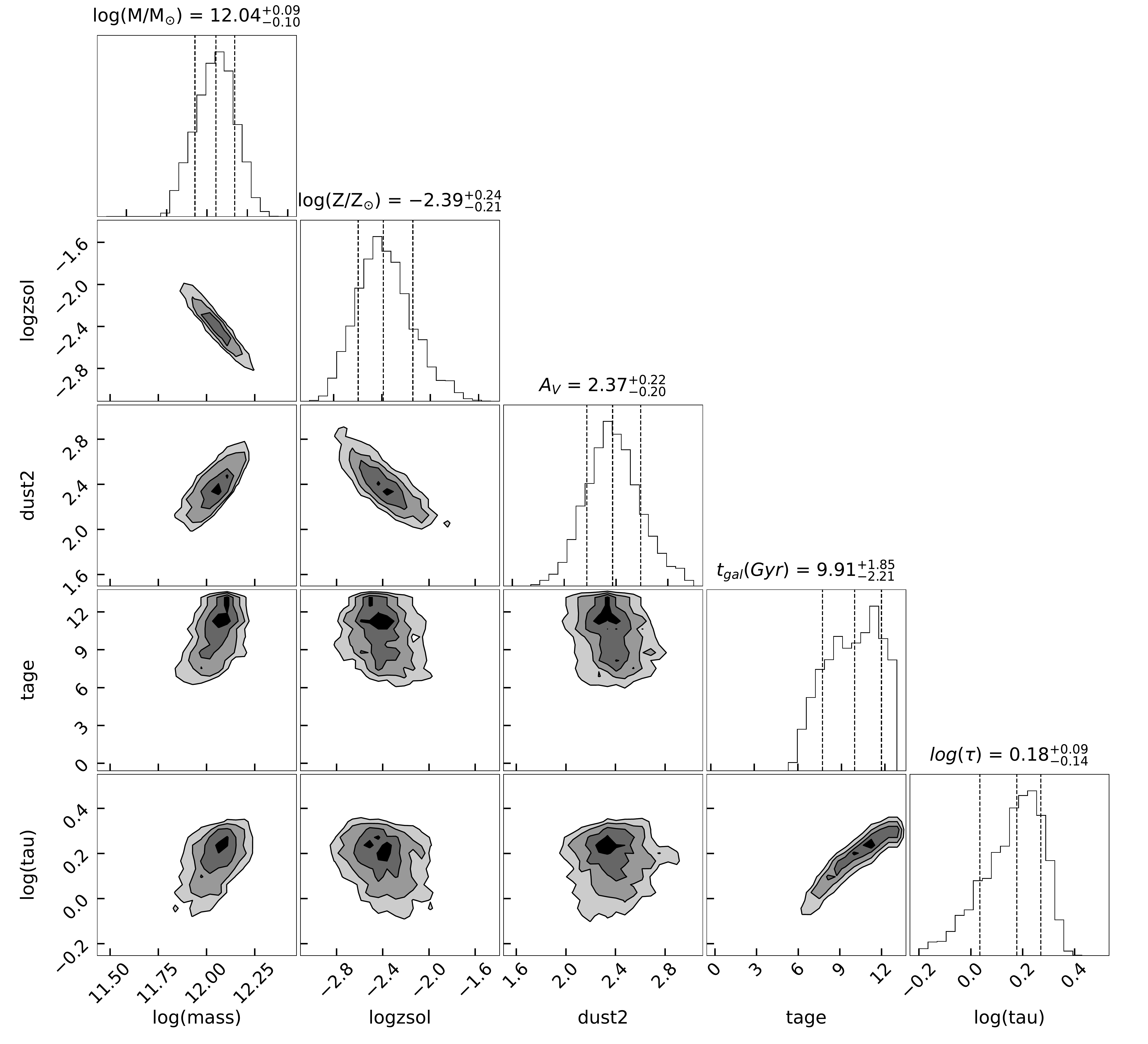}
\includegraphics[width=.9\columnwidth]{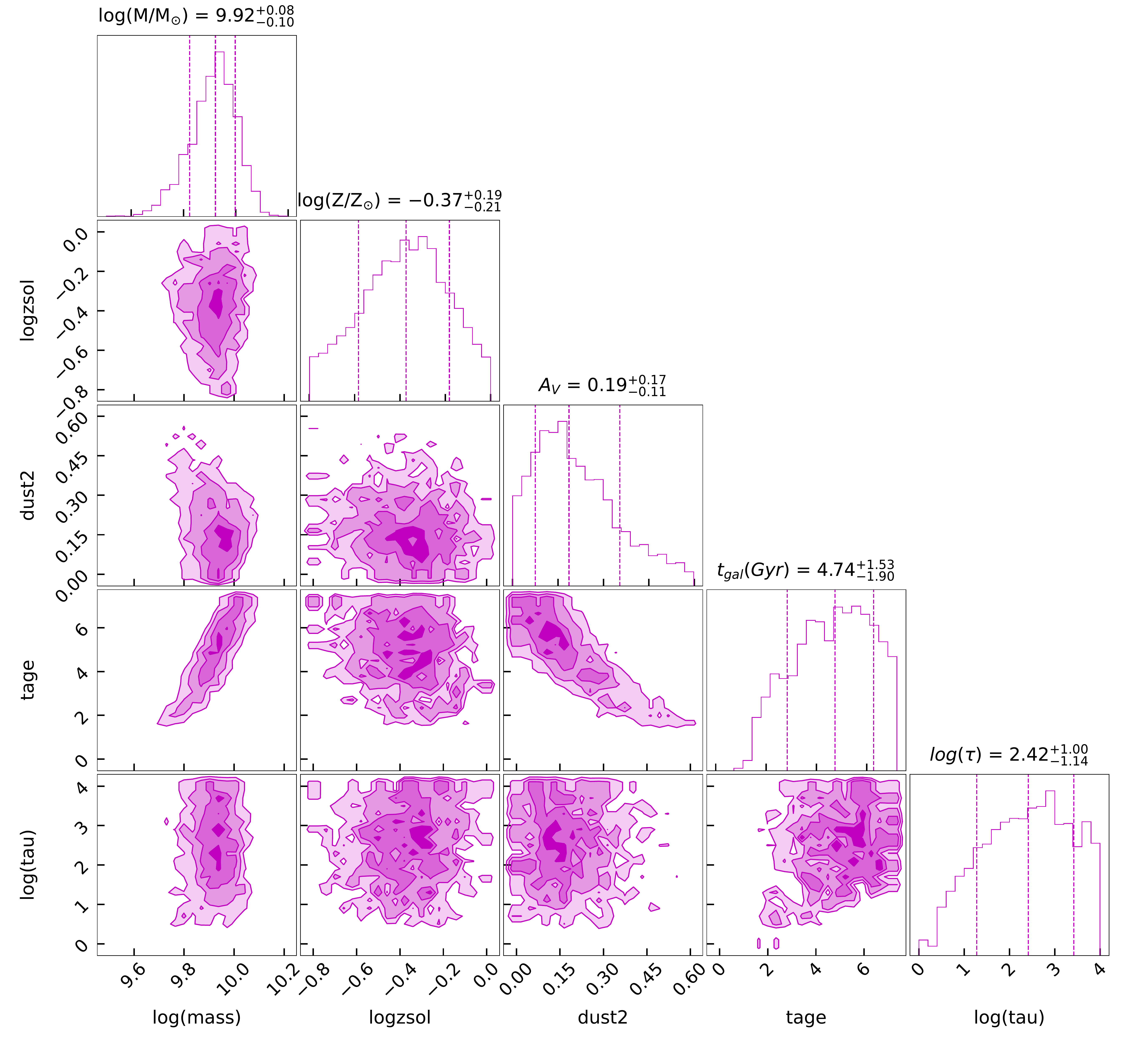}
\caption{The posterior distributions for the SED model parameters of GRB 030329 (cyan), GRB 130603B (blue), GRB 140102A (green), GRB 190829A (black), and GRB 200826A (magenta) obtained using nested sampling via \sw{dynesty} using \sw{Prospector} software.}
\label{corner}
\end{figure*}

\begin{table*}[t]
\caption{Log of the host galaxy panchromatic observations of our sample taken with 3.6m DOT and those reported in the literature. The magnitude shown with star markers is in the Vega system. NOT, CAHA, GTC, and LDT denotes the Nordic Optical Telescope, Centro Astronómico Hispano-Alemán, Gran Telescopio Canarias, and Lowell Discovery Telescope, respectively.}
\begin{center}
\begin{tabular}{c c c c c c}
\hline
\bf Date & \bf Exposure (s) & \bf Magnitude (AB) & \bf Filter & \bf Telescope & \bf References \\ \hline

\multicolumn{6}{c}{\textbf{GRB 030329}} \\ \hline 
24.03.2004 & 5x900 & $23.45 \pm 0.10$ & U & 2.56m NOT & \cite{2005AA...444..711G} \\
24.03.2004 & 3x600 & $23.26 \pm 0.07$ & B & 2.56m NOT & \cite{2005AA...444..711G} \\
05.01.2004& 113x60 & $22.42 \pm 0.16$ & J & 3.5m CAHA & \cite{2005AA...444..711G} \\
06.01.2004& 109x60 & $22.55 \pm 0.24$ & H & 3.5m CAHA & \cite{2005AA...444..711G} \\
07.01.2004& 99x60 & $>21.57 $ & K & 3.5m CAHA & \cite{2005AA...444..711G} \\ 
\bf 23.03.2017&\bf 3x600 & $\bf 22.81 \pm 0.13$ &\bf V &\bf 3.6m DOT &\bf Present work \\
\bf 23.03.2017&\bf 4x600 & $\bf 22.83 \pm 0.07$ &\bf R &\bf 3.6m DOT &\bf Present work \\ \hline

\multicolumn{6}{c}{\textbf{GRB 130603B}} \\ \hline 
05.06.2013 & 5x300 & $20.69 \pm 0.15$ & I & 1.5m OSN & \cite{2019MNRAS.485.5294P} \\
22.06.2013 & 15x60 & $19.69 \pm 0.13$ & K & 3.5m CAHA & \cite{2019MNRAS.485.5294P} \\
22.06.2013 & 15x60 & $20.06 \pm 0.09$ & J & 3.5m CAHA & \cite{2019MNRAS.485.5294P} \\
22.06.2013 & 15x60 & $19.68 \pm 0.13$ & H & 3.5m CAHA & \cite{2019MNRAS.485.5294P} \\
22.06.2013 & 15x60 & $20.11 \pm 0.07$ & Z & 3.5m CAHA & \cite{2019MNRAS.485.5294P} \\
05.07.2013 & 4x50 & $22.01 \pm 0.03$ & g & 10.4m GTC & \cite{2019MNRAS.485.5294P} \\
05.07.2013 & 4x50 & $20.97 \pm 0.01$ & r & 10.4m GTC & \cite{2019MNRAS.485.5294P} \\
05.07.2013 & 4x50 & $20.65 \pm 0.02$ & i & 10.4m GTC & \cite{2019MNRAS.485.5294P} \\
\bf 23.03.2017 &\bf 2x300 &\bf $22.13 \pm 0.05$ &\bf B &\bf 3.6m DOT & \cite{2019MNRAS.485.5294P} \\
\bf 23.03.2017 &\bf 2x300 &\bf $20.72 \pm 0.02$ &\bf R  &\bf 3.6m DOT  & \cite{2019MNRAS.485.5294P}  \\ \hline

\multicolumn{6}{c}{\textbf{GRB 140102A}} \\ \hline

13.05.2014   &   59x65.0   & $   21.18 \pm   0.26 $  &  H  & CAHA & \cite{2021MNRAS.505.4086G}\\
18.07.2017   &   7x120.0   & $   25.13 \pm   0.16 $  &  g  & 10.4m GTC & \cite{2021MNRAS.505.4086G}\\
18.07.2017   &   7x90.0    & $   24.47 \pm   0.13 $  &  r  & 10.4m GTC & \cite{2021MNRAS.505.4086G}\\
 18.07.2017   &   7x90.0    & $   24.17 \pm   0.13 $  &  i  & 10.4m GTC & \cite{2021MNRAS.505.4086G}\\
 18.07.2017   &   6x60.0    & $   23.88 \pm   0.18 $  &  z  & 10.4m GTC & \cite{2021MNRAS.505.4086G}\\
\bf 16.01.2021   & \bf 3x300.0, 2x900.0   & \bf  $\geq$ 24.10  &\bf  R  &\bf  3.6m DOT& \cite{2021MNRAS.505.4086G} \\ \hline

\multicolumn{6}{c}{\textbf{GRB 190829A}} \\ \hline 

26.09.2000 & --&$18.64 \pm 0.05$ &u & SDSS & \cite{2005AJ....129.1755A} \\
26.09.2000 & --& $16.686 \pm 0.005$ & g& SDSS & \cite{2005AJ....129.1755A}\\
26.09.2000 & --&$15.729 \pm 0.004$ & r& SDSS & \cite{2005AJ....129.1755A}\\
26.09.2000 &-- &$15.229 \pm 0.004$ & i& SDSS & \cite{2005AJ....129.1755A}\\
26.09.2000 & --&$14.872 \pm 0.007$ & z& SDSS & \cite{2005AJ....129.1755A}\\
12.12.2000&-- & $13.798 \pm 0.096^{*}$& J & 2MASS &\cite{2006AJ....131.1163S} \\
12.12.2000&-- & $12.877 \pm 0.092^{*}$& H & 2MASS &\cite{2006AJ....131.1163S} \\
12.12.2000&-- & $12.320 \pm 0.104^{*}$& K & 2MASS &\cite{2006AJ....131.1163S} \\
\bf 03.10.2020 &\bf 2x300 & $\bf 17.05 \pm 0.05^{*}$ &\bf B &\bf 3.6m DOT &\bf Present work \\
\bf 03.10.2020 &\bf 2x200 & $\bf 15.75 \pm 0.02^{*}$ &\bf R &\bf 3.6m DOT &\bf Present work \\
\bf 03.10.2020 &\bf 2x200 & $\bf 15.26 \pm 0.03^{*}$ &\bf I &\bf 3.6m DOT &\bf Present work \\ \hline 

\multicolumn{6}{c}{\textbf{GRB 200826A}} \\ \hline 

28.08.2020 & 3600 &  $21.11 \pm  0.16^{*}$  & J & Palomar Hale 200-in (P200) & \cite{2021NatAs.tmp..142A} \\
13.09.2020 & 5x180 & $23.45 \pm 0.24$ & u & 4.3m LDT & \cite{2021NatAs.tmp..142A} \\
13.09.2020 & 4x180 & $23.36 \pm 0.05$ & g & 4.3m LDT & \cite{2021NatAs.tmp..142A} \\
13.09.2020 & 10x150 & $22.86 \pm 0.18$ & r & 4.3m LDT & \cite{2021NatAs.tmp..142A} \\
13.09.2020 & 6x180 & $22.13 \pm 0.05$ & z & 4.3m LDT & \cite{2021NatAs.tmp..142A} \\
\bf 04.11.2020 &\bf 12x300 & $\bf 22.71 \pm 0.10$ &\bf i &\bf 3.6m DOT &\bf Present work \\
\bf 04.11.2020 &\bf 2100 &\bf $>$ 20.56$^{*}$ &\bf J &\bf 3.6m DOT &\bf Present work \\
\bf 11.11.2020 &\bf 2100 &\bf $>$ 19.55$^{*}$ &\bf K &\bf 3.6m DOT &\bf Present work \\ \hline 

\hline
\vspace{-1em}
\end{tabular}
\end{center}
\label{tab:observationslog}
\end{table*}

\section*{Acknowledgements}

We thank the anonymous referee for providing positive and constructive comments to improve the manuscript. RG, and SBP acknowledge BRICS grant {DST/IMRCD/BRICS/PilotCall1/ProFCheap/2017(G)} for the financial support. RG and SBP acknowledge the financial support of ISRO under AstroSat archival Data utilization program (DS$\_$2B-13013(2)/1/2021-Sec.2). AA acknowledges funds and assistance provided by the Council of Scientific \& Industrial Research (CSIR), India with file no. 09/948(0003)/2020-EMR-I. AJCT and SBP acknowledge support from the Spanish Ministry Project PID2020-118491GB-I00. AJCT also acknowledges Junta de Andaluc\'ia Project P20$\_$01068 and the "Center of Excellence Severo Ochoa" award for the Instituto de Astrofísica de Andalucía (SEV-2017-0709)". This research is based on observations obtained at the 3.6m Devasthal Optical Telescope (DOT) which is a National Facility run and managed by Aryabhatta Research Institute of Observational Sciences (ARIES), an autonomous Institute under Department of Science and Technology, Government of India. RG and SBP thank Dr. Youdong Hu. for sharing the data files used to show Fig. \ref{host_sfr} in this paper. RG also thanks Ms. Dimple for helping with \sw{Prospector}.

\end{document}